\documentclass[ngerman,american,aps,pra,reprint, superscriptaddress]{revtex4-1}
\usepackage[T1]{fontenc}
\usepackage[latin9]{inputenc}
\usepackage{color}
\usepackage{babel}
\usepackage{amsmath}
\usepackage{amssymb}
\usepackage{graphicx}
\usepackage[unicode=true,pdfusetitle,
 bookmarks=true,bookmarksnumbered=false,bookmarksopen=false,
 breaklinks=false,pdfborder={0 0 0},backref=false,colorlinks=true]
 {hyperref}

\makeatletter
\usepackage{pxfonts}
\usepackage{braket}

\makeatother

\begin{document}

\title{Entanglement distribution and quantum discord}

\author{Alexander Streltsov}

\email{streltsov.physics@gmail.com}

\affiliation{Dahlem Center for Complex Quantum Systems, Freie Universität Berlin,
D-14195 Berlin, Germany}

\author{Hermann Kampermann}

\affiliation{Institut für Theoretische Physik III, Heinrich-Heine-Universität
Düsseldorf, D-40225 Düsseldorf, Germany}

\author{Dagmar Bruß}

\affiliation{Institut für Theoretische Physik III, Heinrich-Heine-Universität
Düsseldorf, D-40225 Düsseldorf, Germany}
\begin{abstract}
Establishing entanglement between distant parties is one of the most
important problems of quantum technology, since long-distance entanglement
is an essential part of such fundamental tasks as quantum cryptography
or quantum teleportation. In this lecture we review basic properties
of entanglement and quantum discord, and discuss recent results on
entanglement distribution and the role of quantum discord therein.
We also review entanglement distribution with separable states, and
discuss important problems which still remain open. One such open
problem is a possible advantage of indirect entanglement distribution,
when compared to direct distribution protocols. 
\end{abstract}
\maketitle

\section{Introduction}

This lecture presents an overview of the task of establishing entanglement
between two distant parties (Alice and Bob) and its connection to
quantum discord \cite{Ollivier2001,Henderson2001,Modi2012,Streltsov2015b,Adesso2016}.
Surprisingly, it is possible for the two parties to perform this task
successfully by exchanging an ancilla which has never been entangled
with Alice and Bob. This puzzling quantum protocol was already suggested
in \cite{Cubitt2003}, but a thorough study \cite{Streltsov2012,Chuan2012,Kay2012,Streltsov2014,Streltsov2015,Zuppardo2016}
and experimental verification \cite{Fedrizzi2013,Vollmer2013,Peuntinger2013}
(see also \cite{Silberhorn2013}) had to wait for almost ten years
until recently, when interest in general quantum correlations arose
and led to insights about their role in the entanglement distribution
protocol.

A composite quantum system does not need to be in a product state
for the subsystems, but it can also occur as a superposition of product
states, or as a mixture of such superpositions. This feature does
not exist in the classical world, and a state exhibiting it is called
entangled. In general, a state is said to contain entanglement if
it cannot be written as a mixture of projectors onto product states.
It is said to contain quantum correlations, if it cannot be written
as a mixture of projectors onto product states with local orthogonality
properties. And it is said to contain correlations (classical or quantum),
if it cannot be written as a product state.

Let us formalize these notions. In the following definitions we will
consider for simplicity only bipartite quantum systems (with superscripts
$A$ and $B$ for Alice and Bob, respectively); the generalization
to composite quantum systems with more than two subsystems is straightforward.
Let us denote by $\{\ket{e_{i}}\}$ a complete set of orthogonal basis
states (which could also be interpreted as classical states), i.e.
$\braket{e_{i}|e_{j}}=\delta_{ij}$, while Greek letters indicate
quantum states which are not necessarily orthogonal, i.e., for the
ensemble $\{\ket{\psi_{i}}\}$ in general $\braket{\psi_{i}|\psi_{j}}\neq\delta_{ij}$
holds. 

A \textsl{separable} state $\rho_{\mathrm{sep}}$ can be written as~\cite{Werner1989}
\begin{equation}
\rho_{\mathrm{sep}}^{AB}=\sum_{i,j}p_{ij}\ket{\psi_{i}}\!\bra{\psi_{i}}^{A}\otimes\ket{\phi_{j}}\!\bra{\phi_{j}}^{B},\label{eq:sep}
\end{equation}
where $p_{ij}$ are probabilities with $\sum_{i,j}p_{ij}=1$. The
set of all separable states will be denoted by $\mathcal{S}$. Any
separable state can be produced with local operations and classical
communication (LOCC). An \textsl{entangled} state cannot be written
as in Eq.~(\ref{eq:sep}). In order to produce an entangled state,
a non-local operation is needed. In Section \ref{sec:entanglement}
we will review different ways to quantify the amount of entanglement
in a given state.

A state is called \emph{classically correlated} (CC) if it can be
written as \cite{Piani2008}
\begin{equation}
\rho_{\mathrm{cc}}^{AB}=\sum_{i,j}p_{ij}\ket{e_{i}}\!\bra{e_{i}}^{A}\otimes\ket{e_{j}}\!\bra{e_{j}}^{B},\label{eq:CC}
\end{equation}
 with $\braket{e_{i}|e_{j}}=\delta_{ij}$. Measuring $\rho_{\mathrm{cc}}$
in the local bases $\{\ket{e_{i}}^{A}\}$ and $\{\ket{e_{j}}^{B}\}$
does not change the state, i.e., 
\begin{equation}
\Pi^{A}\otimes\Pi^{B}[\rho_{\mathrm{cc}}^{AB}]=\rho_{\mathrm{cc}}^{AB},
\end{equation}
where the von Neumann measurement $\Pi$ is defined as 
\begin{equation}
\Pi[\sigma]=\sum_{i}\ket{e_{i}}\!\bra{e_{i}}\sigma\ket{e_{i}}\!\bra{e_{i}}.\label{eq:measurement}
\end{equation}
The set of all classically correlated states will be denoted by $\mathcal{CC}$.
A \textsl{quantum correlated} state cannot be written as in Eq.~(\ref{eq:CC}).
The eigenbasis of a quantum correlated state is not a product basis
with the property that the sets of local states are orthogonal ensembles. 

It is also possible to combine the aforementioned frameworks of separability
and classicality, thus arriving at \emph{classical-quantum} (CQ) states
\cite{Piani2008}: 
\begin{equation}
\rho_{\mathrm{cq}}^{AB}=\sum_{i,j}p_{ij}\ket{e_{i}}\bra{e_{i}}^{A}\otimes\ket{\psi_{j}}\bra{\psi_{j}}^{B}.\label{eq:CQ}
\end{equation}
The set of all classical-quantum states will be denoted by $\mathcal{CQ}$.
For any CQ state, there exists a local von Neumann measurement on
the subsystem $A$ which leaves the state unchanged, i.e., 
\begin{equation}
\Pi^{A}\otimes\openone^{B}[\rho_{\mathrm{cq}}^{AB}]=\rho_{\mathrm{cq}}^{AB},
\end{equation}
where the von Neumann measurement $\Pi$ is given in Eq.~(\ref{eq:measurement}).
If a state cannot be written as in Eq.~(\ref{eq:CQ}), we say that
the state has nonzero \emph{quantum discord} with respect to the subsystem
$A$. Measuring a state with nonzero discord in any orthogonal basis
on the subsystem $A$ necessarily changes the state. In Section~\ref{sec:discord}
we will present different ways to quantify the amount of discord in
a given state.

\begin{figure}
\includegraphics[width=0.9\columnwidth]{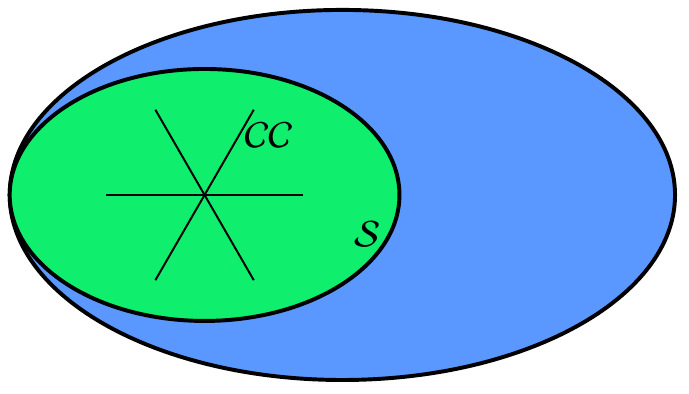} \caption{\selectlanguage{ngerman}%
\label{fig:statespace}\foreignlanguage{american}{State space for
composite quantum systems \cite{Ferraro2010}: classically correlated
states form a connected set $\mathcal{CC}$ of measure zero; separable
states form a convex set $\mathcal{S}$ (green, containing $\mathcal{CC}$).
Quantum correlated states are all states outside of $\mathcal{CC}$,
and entangled states (blue) are all states outside of $\mathcal{S}$.}\selectlanguage{american}%
}
\end{figure}
From the above definitions it is clear that classically correlated
states are a subset of separable states, and that entangled states
are a subset of quantum correlated states. These different types of
states for composite quantum systems therefore form a nested structure
\cite{Ferraro2010} which is sketched in Fig. \ref{fig:statespace}.
Note that separable states form a convex set, due to their definition
in Eq. (\ref{eq:sep}). However, classically correlated states do
\textsl{not} form a convex set: one can produce a quantum correlated
state by mixing two classically correlated states.

Those states which are not entangled, but nevertheless possess quantum
correlations a la discord, may exhibit puzzling features. They can
be produced via LOCC, but nevertheless they carry quantum properties.
Namely, in order to produce them one has to create quantumness in
the form of non-orthogonality. This makes them a potential resource
for quantum information processing protocols. Counterintuitively,
even though they do not carry entanglement, they may be used for the
distribution of entanglement, as we will see below.

The structure of this lecture is as follows: in Section \ref{sec:entanglement}
we review different measures of quantum entanglement, discord quantifiers
are reviewed in Section \ref{sec:discord}. In Section~\ref{sec:entanglement-distribution}
we review recent results on entanglement distribution and discuss
the role of quantum discord therein. Conclusions in Section \ref{sec:conclusions}
complete our lecture.

\section{\label{sec:entanglement}Quantum entanglement}

Here, we will review different entanglement measures, mainly focusing
on measures which are used in this lecture. More detailed reviews,
also containing other entanglement measures, can be found in \cite{Bruss2002,Plenio2007,Horodecki2009}.
In general, we require that a measure of entanglement $\mathcal{E}$
fulfills the following two properties \cite{Vedral1997,Vedral1998}:
\begin{itemize}
\item Nonnegativity: $\mathcal{E}(\rho)\geq0$ for all states $\rho$ with
equality for all separable states \footnote{Note that some entanglement measures (like distillable entanglement
and logarithmic negativity) also vanish for some entangled states.},
\item Monotonicity: $\mathcal{E}(\Lambda[\rho])\leq\mathcal{E}(\rho)$ for
any LOCC operation $\Lambda$.
\end{itemize}
Many entanglement measures also have additional properties such as
strong monotonicity in the sense that entanglement does not increase
on average under selective LOCC operations \cite{Vedral1997,Vedral1998}:
$\sum_{i}p_{i}\mathcal{E}(\sigma_{i})\leq\mathcal{E}(\rho)$, where
the states $\sigma_{i}$ are obtained from the state $\rho$ by the
means of LOCC with the corresponding probabilities $p_{i}$. Moreover,
many entanglement measures are also convex in the state, i.e., $\mathcal{E}(\sum_{i}p_{i}\rho_{i})\leq\sum_{i}p_{i}\mathcal{E}(\rho_{i})$
\cite{Vedral1997,Vedral1998}.

From now on we will focus on the bipartite scenario with two parties
$A$ and $B$ of the same dimension $d$. In this case, any entanglement
measure is maximal on states of the form 
\begin{equation}
\ket{\phi_{d}^{+}}=\frac{1}{\sqrt{d}}\sum_{i=0}^{d-1}\ket{ii},
\end{equation}
since from this state any quantum state can be created via LOCC operations
\cite{Horodecki2009}. Of particular importance is the two-qubit \emph{singlet}
state $(\ket{01}-\ket{10})/\sqrt{2}$, which can be obtained from
the state $\ket{\phi_{2}^{+}}$ via local unitaries. In entanglement
theory local unitaries do not change the properties of a state, and
thus we will refer to the state $\ket{\phi_{2}^{+}}$ as a singlet.

Operational measures of entanglement are \emph{distillable entanglement}
and \emph{entanglement cost}. Distillable entanglement quantifies
the maximal rate for extracting singlets from a state via LOCC operations
\cite{Plenio2007,Horodecki2009}:
\begin{equation}
\mathcal{E}_{\mathrm{d}}(\rho)=\sup\left\{ R:\lim_{n\rightarrow\infty}\left(\inf_{\Lambda}\left\Vert \Lambda\left[\rho^{\otimes n}\right]-\left(\phi_{2}^{+}\right)^{\otimes nR}\right\Vert _{1}\right)=0\right\} ,\label{eq:Ed}
\end{equation}
where $||M||_{1}=\mathrm{Tr}\sqrt{M^{\dagger}M}$ is the trace norm,
$\phi_{2}^{+}$ is the projector onto the state $\ket{\phi_{2}^{+}}$
\footnote{For a general pure state $\ket{\psi}$ we will denote the corresponding
projector by $\psi$, i.e., $\psi=\ket{\psi}\!\bra{\psi}$.}, and the infimum is performed over all LOCC operations $\Lambda$.
Entanglement cost on the other hand quantifies the minimal singlet
rate required for creating a state via LOCC operations~\cite{Plenio2007,Horodecki2009}:
\begin{equation}
\mathcal{E}_{\mathrm{c}}(\rho)=\inf\left\{ R:\lim_{n\rightarrow\infty}\left(\inf_{\Lambda}\left\Vert \Lambda\left[\left(\phi_{2}^{+}\right)^{\otimes nR}\right]-\rho^{\otimes n}\right\Vert _{1}\right)=0\right\} \,.
\end{equation}
For pure states $\ket{\psi}=\ket{\psi}^{AB}$ these two quantities
coincide and are equal to the von Neumann entropy of the reduced state~\cite{Bennett1996b}:
$\mathcal{E}_{\mathrm{d}}(\psi)=\mathcal{E}_{\mathrm{c}}(\psi)=S(\rho^{A})=-\mathrm{Tr}[\rho^{A}\log_{2}\rho^{A}]$.
This implies that the resource theory of entanglement is reversible
for pure states~\cite{Plenio2007,Horodecki2009}. In general, it
holds that $\mathcal{E}_{\mathrm{d}}(\rho)\leq\mathcal{E}_{\mathrm{c}}(\rho)$,
and there exist states which have zero distillable entanglement but
nonzero entanglement cost. This phenomenon is also known as \emph{bound
entanglement} \cite{Horodecki1998}. 

An important family of entanglement measures is obtained by taking
the minimal distance to the set of separable states $\mathcal{S}$~\cite{Vedral1997,Vedral1998}:
\begin{equation}
\mathcal{E}(\rho)=\inf_{\sigma\in\mathcal{S}}D(\rho,\sigma).\label{eq:entanglement}
\end{equation}
Here, $D(\rho,\sigma)$ can be an arbitrary functional which is nonnegative
and monotonic under quantum operations, i.e., $D(\Lambda[\rho],\Lambda[\sigma])\leq D(\rho,\sigma)$
for any quantum operation $\Lambda$ \footnote{Note that $D(\rho,\sigma)$ does not have to be a distance in the
mathematical sense, since it does not necessarily fulfill the triangle
inequality.}. Examples for such distances are the trace distance $||\rho-\sigma||_{1}/2$,
the infidelity $1-F(\rho,\sigma)$ with fidelity $F(\rho,\sigma)=||\sqrt{\rho},\sqrt{\sigma}||_{1}^{2}$,
and the quantum relative entropy $S(\rho||\sigma)=\mathrm{Tr}[\rho\log_{2}\rho]-\mathrm{Tr}[\rho\log_{2}\sigma]$.
In the latter case, the corresponding measure is known as the \emph{relative
entropy of entanglement} \cite{Vedral1997,Vedral1998}: 
\begin{equation}
\mathcal{E}_{\mathrm{r}}(\rho)=\min_{\sigma\in\mathcal{S}}S(\rho||\sigma).
\end{equation}

The second important family of measures are convex roof measures defined
as \cite{Uhlmann1998} 
\begin{equation}
\mathcal{E}(\rho)=\inf\sum_{i}p_{i}\mathcal{E}(\psi_{i}),
\end{equation}
where the infimum is taken over all pure state decompositions of $\rho=\sum_{i}p_{i}\psi_{i}$.
If for pure states entanglement is defined as the von Neumann entropy
of the reduced state $\mathcal{E}(\psi)=S(\rho^{A})$, the corresponding
convex roof measure is known as the \emph{entanglement of formation}
\cite{Bennett1996}:
\begin{equation}
\mathcal{E}_{\mathrm{f}}(\rho)=\min\sum_{i}p_{i}S\left(\mathrm{Tr}_{A}\left[\psi_{i}\right]\right).
\end{equation}
In general, the relative entropy of entanglement is between the distillable
entanglement and the entanglement of formation \cite{Horodecki2000}:
\begin{equation}
\mathcal{E}_{\mathrm{d}}(\rho)\leq\mathcal{E}_{\mathrm{r}}(\rho)\leq\mathcal{E}_{\mathrm{f}}(\rho).
\end{equation}
Moreover, the regularized entanglement of formation is equal to the
entanglement cost \cite{Hayden2001}: $\mathcal{E}_{\mathrm{c}}(\rho)=\lim_{n\rightarrow\infty}\mathcal{E}_{\mathrm{f}}(\rho^{\otimes n})/n$.
We also mention that the geometric measure of entanglement defined
as 
\begin{equation}
\mathcal{E}_{\mathrm{g}}(\rho)=1-\max_{\sigma\in\mathcal{S}}F(\rho,\sigma)
\end{equation}
 is a distance-based and a convex roof measure simultaneously \cite{Wei2003,Streltsov2010}. 

Another important entanglement measure which will be used in this
lecture is the \emph{logarithmic negativity}. For a a bipartite state
$\rho=\rho^{AB}$ it is defined as \cite{Zyczkowski1998,Vidal2002}
\begin{equation}
\mathcal{E}_{\mathrm{n}}(\rho)=\log_{2}\left\Vert \rho^{T_{A}}\right\Vert _{1}
\end{equation}
with the partial transposition $T_{A}$. The logarithmic negativity
is zero for states which have positive partial transpose, and thus
there exist entangled states which have zero logarithmic negativity~\cite{Horodecki1997}.
Nevertheless, these states cannot be distilled into singlets~\cite{Horodecki1998}.
Interestingly, the logarithmic negativity is not convex~\cite{Plenio2005},
and is related to the entanglement cost under quantum operations preserving
the positivity of the partial transpose~\cite{Audenaert2003}. 

Several entanglement measures discussed above are subadditive, i.e.,
they fulfill the inequality 
\begin{equation}
\mathcal{E}\left(\rho\otimes\sigma\right)\leq\mathcal{E}\left(\rho\right)+\mathcal{E}\left(\sigma\right)\label{eq:subadditivity}
\end{equation}
for any two states $\rho$ and $\sigma$. Examples for subadditive
measures are entanglement cost, entanglement of formation, and relative
entropy of entanglement. The logarithmic negativity is additive, i.e.,
it fulfills Eq.~(\ref{eq:subadditivity}) with equality. It is conjectured
\cite{Shor2001} that the distillable entanglement violates Eq.~(\ref{eq:subadditivity})
.

\section{\label{sec:discord}Quantum discord}

Quantum discord was introduced in \cite{Ollivier2001,Henderson2001}
as a quantifier for correlations different from entanglement. In the
modern language of quantum information theory, quantum discord of
a state $\rho=\rho^{AB}$ can be expressed in the following compact
way \cite{Streltsov2013,Streltsov2015c}: 
\begin{equation}
\delta(\rho)=I(\rho)-\sup_{\Lambda_{\mathrm{eb}}}I(\Lambda_{\mathrm{eb}}\otimes\openone\left[\rho\right]).\label{eq:discord}
\end{equation}
Here, $I(\rho^{AB})=S(\rho^{A})+S(\rho^{B})-S(\rho^{AB})$ is the
quantum mutual information and the supremum is performed over all
entanglement breaking channels $\Lambda_{\mathrm{eb}}$~\footnote{An entanglement breaking channel $\Lambda_{\mathrm{eb}}$ has the
property that $\Lambda_{\mathrm{eb}}\otimes\openone[\rho]$ is not
entangled for any bipartite input state $\rho$. We refer to~\cite{Horodecki2003}
for more details.}. Quantum discord vanishes on CQ-states and is larger than zero otherwise
\cite{Datta2010}. The quantity $I(\rho)-\delta(\rho)$ was initially
introduced in \cite{Henderson2001} as a measure of classical correlations.
Interestingly, quantum discord is closely related to the entanglement
of formation via the Koashi-Winter relation \cite{Koashi2004,Fanchini2011}:
\begin{equation}
\delta(\rho^{AB})=E_{\mathrm{f}}(\rho^{BC})-S(\rho^{AB})+S(\rho^{A}),\label{eq:Koashi-Winter}
\end{equation}
where the total state $\rho^{ABC}$ is pure \footnote{Interestingly, Eq.~(\ref{eq:Koashi-Winter}) implies that a simple
formula for quantum discord for all quantum states is out of reach,
since such an expression would also allow for an exact evaluation
of entanglement of formation. Nevertheless, analytical progress on
the evaluation of discord for particular families of states has been
presented in~\cite{Ali2010,Ali2010b,Girolami2011}.}.

Similar as for entanglement, we can define distance-based measures
of discord~\footnote{Many authors also consider the minimal distance to the set of CQ states,
i.e., $\Delta(\rho)=\inf_{\sigma\in\mathcal{CQ}}D(\rho,\sigma)$.
We note that $\Delta$ and $\mathcal{D}$ coincide for the quantum
relative entropy, and $\Delta(\rho)\leq\mathcal{D}(\rho)$ in general
\cite{Modi2012}.}: 
\begin{equation}
\mathcal{D}(\rho)=\inf_{\Pi}D(\rho,\Pi\otimes\openone[\rho]),\label{eq:discord-2}
\end{equation}
where the infimum is performed over all local von Neumann measurements
$\Pi$ and $D(\rho,\sigma)$ is a suitable distance between $\rho$
and $\sigma$, such as the relative entropy. In the latter case, the
corresponding quantity is called \emph{relative entropy of discord}~\cite{Modi2010}:
\begin{equation}
\mathcal{D}_{\mathrm{r}}(\rho)=\min_{\Pi}S(\rho||\Pi\otimes\openone[\rho]),
\end{equation}
and has also been studied earlier in the context of thermodynamics
\cite{Oppenheim2002,Horodecki2005}. If the distance is chosen to
be the squared Hilbert-Schmidt distance $\mathrm{Tr}(\rho-\sigma)^{2}$,
the corresponding measure is known as the \emph{geometric discord~}\cite{Dakic2010,Luo2010}.
Interestingly, the geometric discord can increase under local operations
on any of the subsystems \cite{Piani2012}. It was also shown to play
a role for remote state preparation~\cite{Dakic2012}.

The role of quantum discord in quantum information theory has been
studied extensively in the last years~\cite{Modi2012,Streltsov2015b}.
Several alternative quantifiers of discord have been presented~\cite{Adesso2016},
and criteria for good discord measures have also been discussed~\cite{Brodutch2011}.
As an important example, we mention the \emph{interferometric power}~\cite{Girolami2014},
which is a computable measure of discord and a figure of merit in
the task of phase estimation with bipartite states. Further results
on the role of quantum discord in quantum metrology have been presented
in~\cite{Modi2011,Gilolami2013}. The relation between quantum discord
and entanglement creation in the quantum measurement process has also
been investigated, both theoretically~\cite{Streltsov2011b,Piani2011}
and experimentally~\cite{Adesso2014}. Monogamy of quantum discord~\cite{Streltsov2012b,Dhar2016}
and its behavior under local noise~\cite{Streltsov2011} and non-Markovian
dynamics~\cite{Fanchini2010} have also been studied. Experimentally
friendly measures of discord were presented in~\cite{Auccaise2011,Girolami2012},
and the possibility of local detection of discord has been reported
in~\cite{Gessner2014}. As we will see in the next section, quantum
discord also plays an important role for entanglement distribution
\cite{Streltsov2012,Chuan2012}.

\section{\label{sec:entanglement-distribution}Entanglement distribution}

\begin{figure}
\includegraphics[width=1\columnwidth]{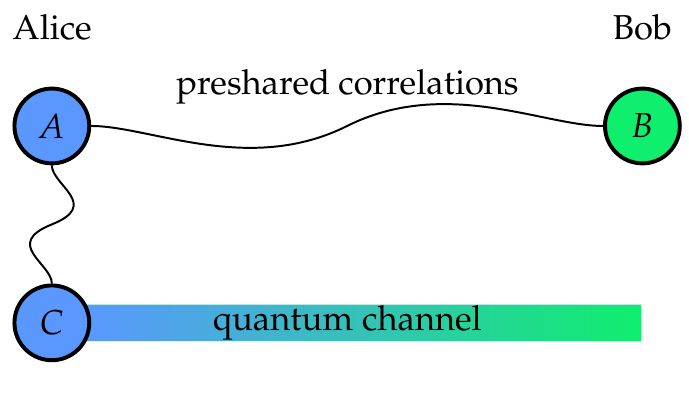}

\caption{\label{fig:entanglement_distribution} Indirect entanglement distribution.
Alice and Bob initially share the state $\rho=\rho^{ABC}$, where
Alice holds the particles $A$ and $C$, and Bob holds the particle
$B$. Entanglement distribution is achieved by sending the particle
$C$ from Alice to Bob via a (possibly noisy) quantum channel. The
figure is taken from \cite{Streltsov2015}.}

\end{figure}

In the following discussion we will distinguish between \emph{direct}
and \emph{indirect} entanglement distribution between two parties
(Alice and Bob) \cite{Streltsov2015,Zuppardo2016}. Direct entanglement
distribution is achieved if Alice prepares two particles in an entangled
state $\rho$ and sends one of them to Bob. The amount of distributed
entanglement is then given by $\mathcal{E}(\openone\otimes\Lambda[\rho])$,
where $\Lambda$ describes the corresponding quantum channel, and
$\mathcal{E}$ is a suitable entanglement measure.

Indirect entanglement distribution is a more general scenario where
Alice and Bob already share correlations initially. In this case the
total initial state is a tripartite state $\rho=\rho^{ABC}$, where
Alice is in possession of the particles $A$ and $C$, and the particle
$B$ is in Bob's hands. Entanglement distribution is then achieved
by sending the particle $C$ from Alice to Bob, see Fig. \ref{fig:entanglement_distribution}.
The amount of distributed entanglement is then given by $\mathcal{E}^{A|BC}(\openone^{AB}\otimes\Lambda_{C}[\rho])-\mathcal{E}^{AC|B}(\rho)$.
In the following, we will discuss recent results on these two types
of entanglement distribution \cite{Streltsov2015,Zuppardo2016}.

\subsection{Direct entanglement distribution}

What is the maximal amount of entanglement that can be directly distributed
via a given quantum channel~$\Lambda$? For answering this question,
we first introduce the corresponding figure of merit: 
\begin{equation}
\mathcal{E}_{\mathrm{direct}}(\Lambda)=\sup_{\sigma}\mathcal{E}\left(\openone\otimes\Lambda[\sigma]\right).\label{eq:E_no_correlations}
\end{equation}
In general, the supremum is performed over all bipartite quantum states
$\sigma$. However, if the entanglement quantifier $\mathcal{E}$
is convex, we can restrict ourselves to pure states. 

If the distribution channel is noiseless, i.e., $\Lambda=\openone$,
then Eq.~(\ref{eq:E_no_correlations}) reduces to 
\begin{equation}
\mathcal{E}_{\mathrm{direct}}(\openone)=\mathcal{E}(\phi_{d}^{+}),
\end{equation}
where $d$ is the dimension of the carrier particle. It is tempting
to believe that this also extends to noisy channels, i.e., that for
any noisy channel  the optimal performance is achieved by sending
one half of a maximally entangled state. Quite surprisingly, this
procedure is not optimal in general \cite{Ziman2007,Pal2014,Streltsov2015}.
In particular, for any convex entanglement measure $\mathcal{E}$
there exists a noisy channel $\Lambda$ and a bipartite state $\rho$
such that~\cite{Ziman2007} 
\begin{equation}
\mathcal{E}(\openone\otimes\Lambda[\rho])>\mathcal{E}(\openone\otimes\Lambda[\phi_{d}^{+}]).
\end{equation}
Even more, if entanglement is quantified via the logarithmic negativity,
then states with arbitrary little entanglement can outperform maximally
entangled states for some noisy channels \cite{Streltsov2015}. Nevertheless,
maximally entangled states are still optimal in various scenarios,
e.g. if the carrier particle is a qubit and entanglement quantifier
is the entanglement of formation or the geometric entanglement \cite{Streltsov2015}.
If the distribution channel is a single-qubit Pauli channel, i.e.,
\begin{equation}
\Lambda_{\mathrm{p}}[\rho]=\sum_{i=0}^{3}p_{i}\sigma_{i}\rho\sigma_{i},\label{eq:pauli}
\end{equation}
where $\sigma_{i}$ are Pauli matrices with $\sigma_{0}=\openone$,
then maximally entangled states are optimal for entanglement distribution,
regardless of the particular entanglement measure~\cite{Streltsov2015}:
\begin{equation}
\mathcal{E}_{\mathrm{direct}}(\Lambda_{\mathrm{p}})=\mathcal{E}(\openone\otimes\Lambda_{\mathrm{p}}[\phi_{2}^{+}]).
\end{equation}
This result also holds if entanglement distribution is performed via
a combination of (possibly different) Pauli channels, also in this
case sending one half of a maximally entangled state is the best strategy.
Finally, if entanglement is quantified via the logarithmic negativity,
maximally entangled states are optimal for all unital single-qubit
channels \footnote{The authors of \cite{Pal2014} proved this statement for negativity
$\mathcal{N}$, which is related to the logarithmic negativity as
$\mathcal{E}_{\mathrm{n}}=\log_{2}(2\mathcal{N}+1)$. Since $\mathcal{N}$
is a nondecreasing function of $\mathcal{E}_{\mathrm{n}}$, it follows
that the statement is also true for the logarithmic negativity.}.

This completes our discussion on direct entanglement distribution,
and we will present the more general scenario in the following.

\subsection{Indirect entanglement distribution}

Can Alice and Bob gain an advantage if they share some correlations
initially? To answer this question, we first introduce a figure of
merit for indirect entanglement distribution:
\begin{equation}
\mathcal{E}_{\mathrm{indirect}}(\Lambda)=\sup_{\rho}\left\{ \mathcal{E}^{A|BC}(\openone^{AB}\otimes\Lambda_{C}[\rho])-\mathcal{E}^{AC|B}(\rho)\right\} ,
\end{equation}
where the supremum is taken over all tripartite states $\rho=\rho^{ABC}$.
In particular, we are interested in the question if $\mathcal{E}_{\mathrm{indirect}}$
is larger than $\mathcal{E}_{\mathrm{direct}}$ for some noisy channel
and some entanglement measure.

Note that so far no general answer to this question is known, and
partial results have been presented in~\cite{Streltsov2015,Zuppardo2016}.
In particular, if the channel used for entanglement distribution is
a single-qubit Pauli channel given in Eq.~(\ref{eq:pauli}) and entanglement
is quantified via a subadditive measure $\mathcal{E}$, then indirect
entanglement distribution does not provide any advantage \cite{Streltsov2015}:
\begin{equation}
\mathcal{E}_{\mathrm{indirect}}(\Lambda_{\mathrm{p}})=\mathcal{E}_{\mathrm{direct}}(\Lambda_{\mathrm{p}})=\mathcal{E}(\openone\otimes\Lambda_{\mathrm{p}}[\phi_{2}^{+}]).
\end{equation}
This means that in this case sending one half of a singlet state is
the optimal distribution strategy. This result can be generalized
to the case where entanglement is distributed via a combination of
(possibly different) Pauli channels \cite{Streltsov2015}.

However, not all entanglement measures are subadditive. An important
example is the distillable entanglement $\mathcal{E}_{\mathrm{d}}$
which was defined in Eq.~(\ref{eq:Ed}) and is conjectured~\cite{Shor2001}
to violate subadditivity. Interestingly, if this conjecture is true,
then indirect entanglement distribution provides an advantage for
the distribution of distillable entanglement \cite{Streltsov2015}.

Finally, we note that entanglement breaking channels  cannot be used
for entanglement distribution for any entanglement measure $\mathcal{E}$
\cite{Zuppardo2016}: 
\begin{equation}
\mathcal{E}_{\mathrm{indirect}}(\Lambda_{\mathrm{eb}})=\mathcal{E}_{\mathrm{direct}}(\Lambda_{\mathrm{eb}})=0
\end{equation}
for any entanglement breaking channel $\Lambda_{\mathrm{eb}}$. This
can be seen by noting that any entanglement breaking channel is equivalent
to an LOCC protocol \cite{Horodecki2003}.

\subsection{Entanglement distribution with separable states}

Entanglement can also be distributed by sending a carrier particle
which is not entangled with the rest of the system. In particular,
there exist tripartite states $\rho=\rho^{ABC}$ such that 
\begin{eqnarray}
\mathcal{E}^{AC|B}(\rho)=\mathcal{E}^{AB|C}(\rho)=0, & \,\,\,\,\,\,\, & \mathcal{E}^{A|BC}(\rho)>0.\label{eq:separable}
\end{eqnarray}
The first example for a state fulfilling Eqs.~(\ref{eq:separable})
was presented in \cite{Cubitt2003}, and can be written as
\begin{equation}
\eta=\frac{1}{3}\ket{\Psi_{\mathrm{GHZ}}}\!\bra{\Psi_{\mathrm{GHZ}}}+\sum_{i,j,k=0}^{1}\beta_{ijk}\Pi_{ijk},\label{eq:cubitt}
\end{equation}
with $\ket{\Psi_{\mathrm{GHZ}}}=(\ket{000}+\ket{111})/\sqrt{2}$,
$\Pi_{ijk}=\ket{ijk}\!\bra{ijk}$, and all $\beta_{ijk}$ are zero
apart from $\beta_{001}=\beta_{010}=\beta_{101}=\beta_{110}=1/6$.
These results were extended to Gaussian states in \cite{Mista2008},
and experiments verifying this phenomenon have also been reported
\cite{Fedrizzi2013,Vollmer2013,Peuntinger2013}.

Motivated by this result, Zuppardo \emph{et al}. \cite{Zuppardo2016}
proposed a classification of entanglement distribution protocols.
In particular, a noiseless distribution protocol is called \emph{excessive}
if the amount of distributed entanglement is larger than the amount
of entanglement between the carrier and the rest of the system, i.e.,
\begin{equation}
\mathcal{E}^{A|BC}(\rho)-\mathcal{E}^{AC|B}(\rho)>\mathcal{E}^{AB|C}(\rho).
\end{equation}
Otherwise, the protocol is called nonexcessive. As discussed above,
the state $\eta$ in Eq.~(\ref{eq:cubitt}) gives rise to an excessive
distribution protocol. 

It is natural to ask if such entanglement distribution with separable
states can provide an advantage when compared to scenarios where the
carrier particle is entangled with the rest of the system. In particular,
one might ask if a separable state can show a better performance for
entanglement distribution when compared to maximally entangled states.
This question could be especially relevant if the distribution channel
is noisy. Despite attempts by several authors \cite{Kay2012,Fedrizzi2013},
the question has not yet been settled. 

Finally, we mention that rank two separable states are not useful
for entanglement distribution if entanglement is quantified via logarithmic
negativity~\cite{Streltsov2014}.

\subsection{Role of quantum discord for entanglement distribution}

As was shown in \cite{Streltsov2012,Chuan2012}, the amount of entanglement
that can be distributed via a noiseless channel by using a tripartite
quantum state $\rho=\rho^{ABC}$ is bounded above by the discord between
the carrier particle $C$ and the rest of the system:
\begin{equation}
\mathcal{E}^{A|BC}(\rho)-\mathcal{E}^{AC|B}(\rho)\leq\mathcal{D}^{C|AB}(\rho).\label{eq:Discord-Bound}
\end{equation}
This inequality is true for any distance-based measure of entanglement
and discord given in Eqs.~(\ref{eq:entanglement}) and (\ref{eq:discord-2})
if the corresponding distance does not increase under quantum operations
and fulfills the triangle inequality. Moreover, it is also true for
the relative entropy of entanglement and discord~\cite{Streltsov2012,Chuan2012}. 

The inequality (\ref{eq:Discord-Bound}) immediately implies that
zero-discord states cannot be used for entanglement distribution.
Moreover, this result can also be used to bound the amount of entanglement
in one cut of a tripartite state $\rho=\rho^{ABC}$ in terms of entanglement
and discord in the other cuts~\cite{Streltsov2012,Chuan2012}: 
\begin{equation}
\mathcal{E}^{AC|B}(\rho)+\mathcal{D}^{C|AB}(\rho)\geq\mathcal{E}^{A|BC}(\rho)\geq\mathcal{E}^{AC|B}(\rho)-\mathcal{D}^{C|AB}(\rho).
\end{equation}
For the relative entropy of entanglement and discord, the inequality~(\ref{eq:Discord-Bound})
is saturated for pure states of the form $\ket{\psi}^{AC}\otimes\ket{\phi}^{B}$
and also for the state $\eta$ given in Eq.~(\ref{eq:cubitt}) \cite{Streltsov2012}. 

If the channel used for entanglement distribution is noisy, we get
the following generalized inequality \cite{Streltsov2015}: 
\begin{align}
\mathcal{E}^{A|BC}(\rho')-\mathcal{E}^{AC|B}(\rho) & \leq\min\left\{ \mathcal{D}^{C|AB}(\rho),\mathcal{D}^{C|AB}(\rho')\right\} .
\end{align}
Here, we used the notation $\rho'=\openone^{AB}\otimes\Lambda_{C}[\rho]$,
and $\mathcal{E}$ and $\mathcal{D}$ are any measures of entanglement
and discord which fulfill Eq.~(\ref{eq:Discord-Bound}).

\section{\label{sec:conclusions}Conclusions}

In this lecture we discussed recent results on entanglement distribution
and the role of quantum discord in this task. Despite substantial
progress in recent years, several important questions in this research
field still remain open. In particular, it is still unclear if indirect
entanglement distribution can provide an advantage in comparison to
direct distribution protocols. The question also concerns entanglement
distribution with separable states: also in this case it remains unclear
if such scheme can be more useful than any direct distribution procedure. 

We also mention that studying entanglement distribution in relation
to the resource theory of coherence~\cite{Baumgratz2014,Winter2016,Streltsov2016}
and its extension to distributed scenarios~\cite{Bromley2015,Streltsov2015e,Chitambar2016,Ma2016,Chitambar2016b,Matera2016,Streltsov2015d,Yadin2015}
could potentially shed new light on these questions, and also lead
to new independent results. 
\begin{acknowledgments}
We thank Remigiusz Augusiak, Maciej Demianowicz, Jens Eisert, and
Maciej Lewenstein for discussion. This work was supported by the Alexander
von Humboldt-Foundation, Bundesministerium für Bildung und Forschung,
and Deutsche Forschungsgemeinschaft.
\end{acknowledgments}

\bibliographystyle{apsrev4-1}
\bibliography{literature}

\begin{thebibliography}{89}%
\makeatletter
\providecommand \@ifxundefined [1]{%
 \@ifx{#1\undefined}
}%
\providecommand \@ifnum [1]{%
 \ifnum #1\expandafter \@firstoftwo
 \else \expandafter \@secondoftwo
 \fi
}%
\providecommand \@ifx [1]{%
 \ifx #1\expandafter \@firstoftwo
 \else \expandafter \@secondoftwo
 \fi
}%
\providecommand \natexlab [1]{#1}%
\providecommand \enquote  [1]{``#1''}%
\providecommand \bibnamefont  [1]{#1}%
\providecommand \bibfnamefont [1]{#1}%
\providecommand \citenamefont [1]{#1}%
\providecommand \href@noop [0]{\@secondoftwo}%
\providecommand \href [0]{\begingroup \@sanitize@url \@href}%
\providecommand \@href[1]{\@@startlink{#1}\@@href}%
\providecommand \@@href[1]{\endgroup#1\@@endlink}%
\providecommand \@sanitize@url [0]{\catcode `\\12\catcode `\$12\catcode
  `\&12\catcode `\#12\catcode `\^12\catcode `\_12\catcode `\%12\relax}%
\providecommand \@@startlink[1]{}%
\providecommand \@@endlink[0]{}%
\providecommand \url  [0]{\begingroup\@sanitize@url \@url }%
\providecommand \@url [1]{\endgroup\@href {#1}{\urlprefix }}%
\providecommand \urlprefix  [0]{URL }%
\providecommand \Eprint [0]{\href }%
\providecommand \doibase [0]{http://dx.doi.org/}%
\providecommand \selectlanguage [0]{\@gobble}%
\providecommand \bibinfo  [0]{\@secondoftwo}%
\providecommand \bibfield  [0]{\@secondoftwo}%
\providecommand \translation [1]{[#1]}%
\providecommand \BibitemOpen [0]{}%
\providecommand \bibitemStop [0]{}%
\providecommand \bibitemNoStop [0]{.\EOS\space}%
\providecommand \EOS [0]{\spacefactor3000\relax}%
\providecommand \BibitemShut  [1]{\csname bibitem#1\endcsname}%
\let\auto@bib@innerbib\@empty
%</preamble>
\bibitem [{\citenamefont {Ollivier}\ and\ \citenamefont
  {Zurek}(2001)}]{Ollivier2001}%
  \BibitemOpen
  \bibfield  {author} {\bibinfo {author} {\bibfnamefont {H.}~\bibnamefont
  {Ollivier}}\ and\ \bibinfo {author} {\bibfnamefont {W.~H.}\ \bibnamefont
  {Zurek}},\ }\href {\doibase 10.1103/PhysRevLett.88.017901} {\bibfield
  {journal} {\bibinfo  {journal} {Phys. Rev. Lett.}\ }\textbf {\bibinfo
  {volume} {88}},\ \bibinfo {pages} {017901} (\bibinfo {year}
  {2001})}\BibitemShut {NoStop}%
\bibitem [{\citenamefont {Henderson}\ and\ \citenamefont
  {Vedral}(2001)}]{Henderson2001}%
  \BibitemOpen
  \bibfield  {author} {\bibinfo {author} {\bibfnamefont {L.}~\bibnamefont
  {Henderson}}\ and\ \bibinfo {author} {\bibfnamefont {V.}~\bibnamefont
  {Vedral}},\ }\href {\doibase 10.1088/0305-4470/34/35/315} {\bibfield
  {journal} {\bibinfo  {journal} {J. Phys. A}\ }\textbf {\bibinfo {volume}
  {34}},\ \bibinfo {pages} {6899} (\bibinfo {year} {2001})}\BibitemShut
  {NoStop}%
\bibitem [{\citenamefont {Modi}\ \emph {et~al.}(2012)\citenamefont {Modi},
  \citenamefont {Brodutch}, \citenamefont {Cable}, \citenamefont {Paterek},\
  and\ \citenamefont {Vedral}}]{Modi2012}%
  \BibitemOpen
  \bibfield  {author} {\bibinfo {author} {\bibfnamefont {K.}~\bibnamefont
  {Modi}}, \bibinfo {author} {\bibfnamefont {A.}~\bibnamefont {Brodutch}},
  \bibinfo {author} {\bibfnamefont {H.}~\bibnamefont {Cable}}, \bibinfo
  {author} {\bibfnamefont {T.}~\bibnamefont {Paterek}}, \ and\ \bibinfo
  {author} {\bibfnamefont {V.}~\bibnamefont {Vedral}},\ }\href {\doibase
  10.1103/RevModPhys.84.1655} {\bibfield  {journal} {\bibinfo  {journal} {Rev.
  Mod. Phys.}\ }\textbf {\bibinfo {volume} {84}},\ \bibinfo {pages} {1655}
  (\bibinfo {year} {2012})}\BibitemShut {NoStop}%
\bibitem [{\citenamefont {Streltsov}(2015)}]{Streltsov2015b}%
  \BibitemOpen
  \bibfield  {author} {\bibinfo {author} {\bibfnamefont {A.}~\bibnamefont
  {Streltsov}},\ }\href {\doibase 10.1007/978-3-319-09656-8} {\emph {\bibinfo
  {title} {{Quantum Correlations Beyond Entanglement and their Role in Quantum
  Information Theory}}}}\ (\bibinfo  {publisher} {SpringerBriefs in Physics},\
  \bibinfo {year} {2015})\ \Eprint {http://arxiv.org/abs/arXiv:1411.3208}
  {arXiv:1411.3208} \BibitemShut {NoStop}%
\bibitem [{\citenamefont {{Adesso}}\ \emph {et~al.}(2016)\citenamefont
  {{Adesso}}, \citenamefont {{Bromley}},\ and\ \citenamefont
  {{Cianciaruso}}}]{Adesso2016}%
  \BibitemOpen
  \bibfield  {author} {\bibinfo {author} {\bibfnamefont {G.}~\bibnamefont
  {{Adesso}}}, \bibinfo {author} {\bibfnamefont {T.~R.}\ \bibnamefont
  {{Bromley}}}, \ and\ \bibinfo {author} {\bibfnamefont {M.}~\bibnamefont
  {{Cianciaruso}}},\ }\href@noop {} {} (\bibinfo {year} {2016}),\ \Eprint
  {http://arxiv.org/abs/1605.00806} {arXiv:1605.00806} \BibitemShut {NoStop}%
\bibitem [{\citenamefont {Cubitt}\ \emph {et~al.}(2003)\citenamefont {Cubitt},
  \citenamefont {Verstraete}, \citenamefont {D\"ur},\ and\ \citenamefont
  {Cirac}}]{Cubitt2003}%
  \BibitemOpen
  \bibfield  {author} {\bibinfo {author} {\bibfnamefont {T.~S.}\ \bibnamefont
  {Cubitt}}, \bibinfo {author} {\bibfnamefont {F.}~\bibnamefont {Verstraete}},
  \bibinfo {author} {\bibfnamefont {W.}~\bibnamefont {D\"ur}}, \ and\ \bibinfo
  {author} {\bibfnamefont {J.~I.}\ \bibnamefont {Cirac}},\ }\href {\doibase
  10.1103/PhysRevLett.91.037902} {\bibfield  {journal} {\bibinfo  {journal}
  {Phys. Rev. Lett.}\ }\textbf {\bibinfo {volume} {91}},\ \bibinfo {pages}
  {037902} (\bibinfo {year} {2003})}\BibitemShut {NoStop}%
\bibitem [{\citenamefont {Streltsov}\ \emph
  {et~al.}(2012{\natexlab{a}})\citenamefont {Streltsov}, \citenamefont
  {Kampermann},\ and\ \citenamefont {Bru\ss{}}}]{Streltsov2012}%
  \BibitemOpen
  \bibfield  {author} {\bibinfo {author} {\bibfnamefont {A.}~\bibnamefont
  {Streltsov}}, \bibinfo {author} {\bibfnamefont {H.}~\bibnamefont
  {Kampermann}}, \ and\ \bibinfo {author} {\bibfnamefont {D.}~\bibnamefont
  {Bru\ss{}}},\ }\href {\doibase 10.1103/PhysRevLett.108.250501} {\bibfield
  {journal} {\bibinfo  {journal} {Phys. Rev. Lett.}\ }\textbf {\bibinfo
  {volume} {108}},\ \bibinfo {pages} {250501} (\bibinfo {year}
  {2012}{\natexlab{a}})}\BibitemShut {NoStop}%
\bibitem [{\citenamefont {Chuan}\ \emph {et~al.}(2012)\citenamefont {Chuan},
  \citenamefont {Maillard}, \citenamefont {Modi}, \citenamefont {Paterek},
  \citenamefont {Paternostro},\ and\ \citenamefont {Piani}}]{Chuan2012}%
  \BibitemOpen
  \bibfield  {author} {\bibinfo {author} {\bibfnamefont {T.~K.}\ \bibnamefont
  {Chuan}}, \bibinfo {author} {\bibfnamefont {J.}~\bibnamefont {Maillard}},
  \bibinfo {author} {\bibfnamefont {K.}~\bibnamefont {Modi}}, \bibinfo {author}
  {\bibfnamefont {T.}~\bibnamefont {Paterek}}, \bibinfo {author} {\bibfnamefont
  {M.}~\bibnamefont {Paternostro}}, \ and\ \bibinfo {author} {\bibfnamefont
  {M.}~\bibnamefont {Piani}},\ }\href {\doibase 10.1103/PhysRevLett.109.070501}
  {\bibfield  {journal} {\bibinfo  {journal} {Phys. Rev. Lett.}\ }\textbf
  {\bibinfo {volume} {109}},\ \bibinfo {pages} {070501} (\bibinfo {year}
  {2012})}\BibitemShut {NoStop}%
\bibitem [{\citenamefont {Kay}(2012)}]{Kay2012}%
  \BibitemOpen
  \bibfield  {author} {\bibinfo {author} {\bibfnamefont {A.}~\bibnamefont
  {Kay}},\ }\href {\doibase 10.1103/PhysRevLett.109.080503} {\bibfield
  {journal} {\bibinfo  {journal} {Phys. Rev. Lett.}\ }\textbf {\bibinfo
  {volume} {109}},\ \bibinfo {pages} {080503} (\bibinfo {year}
  {2012})}\BibitemShut {NoStop}%
\bibitem [{\citenamefont {Streltsov}\ \emph {et~al.}(2014)\citenamefont
  {Streltsov}, \citenamefont {Kampermann},\ and\ \citenamefont
  {Bru\ss{}}}]{Streltsov2014}%
  \BibitemOpen
  \bibfield  {author} {\bibinfo {author} {\bibfnamefont {A.}~\bibnamefont
  {Streltsov}}, \bibinfo {author} {\bibfnamefont {H.}~\bibnamefont
  {Kampermann}}, \ and\ \bibinfo {author} {\bibfnamefont {D.}~\bibnamefont
  {Bru\ss{}}},\ }\href {\doibase 10.1103/PhysRevA.90.032323} {\bibfield
  {journal} {\bibinfo  {journal} {Phys. Rev. A}\ }\textbf {\bibinfo {volume}
  {90}},\ \bibinfo {pages} {032323} (\bibinfo {year} {2014})}\BibitemShut
  {NoStop}%
\bibitem [{\citenamefont {Streltsov}\ \emph
  {et~al.}(2015{\natexlab{a}})\citenamefont {Streltsov}, \citenamefont
  {Augusiak}, \citenamefont {Demianowicz},\ and\ \citenamefont
  {Lewenstein}}]{Streltsov2015}%
  \BibitemOpen
  \bibfield  {author} {\bibinfo {author} {\bibfnamefont {A.}~\bibnamefont
  {Streltsov}}, \bibinfo {author} {\bibfnamefont {R.}~\bibnamefont {Augusiak}},
  \bibinfo {author} {\bibfnamefont {M.}~\bibnamefont {Demianowicz}}, \ and\
  \bibinfo {author} {\bibfnamefont {M.}~\bibnamefont {Lewenstein}},\ }\href
  {\doibase 10.1103/PhysRevA.92.012335} {\bibfield  {journal} {\bibinfo
  {journal} {Phys. Rev. A}\ }\textbf {\bibinfo {volume} {92}},\ \bibinfo
  {pages} {012335} (\bibinfo {year} {2015}{\natexlab{a}})}\BibitemShut
  {NoStop}%
\bibitem [{\citenamefont {Zuppardo}\ \emph {et~al.}(2016)\citenamefont
  {Zuppardo}, \citenamefont {Krisnanda}, \citenamefont {Paterek}, \citenamefont
  {Bandyopadhyay}, \citenamefont {Banerjee}, \citenamefont {Deb}, \citenamefont
  {Halder}, \citenamefont {Modi},\ and\ \citenamefont
  {Paternostro}}]{Zuppardo2016}%
  \BibitemOpen
  \bibfield  {author} {\bibinfo {author} {\bibfnamefont {M.}~\bibnamefont
  {Zuppardo}}, \bibinfo {author} {\bibfnamefont {T.}~\bibnamefont {Krisnanda}},
  \bibinfo {author} {\bibfnamefont {T.}~\bibnamefont {Paterek}}, \bibinfo
  {author} {\bibfnamefont {S.}~\bibnamefont {Bandyopadhyay}}, \bibinfo {author}
  {\bibfnamefont {A.}~\bibnamefont {Banerjee}}, \bibinfo {author}
  {\bibfnamefont {P.}~\bibnamefont {Deb}}, \bibinfo {author} {\bibfnamefont
  {S.}~\bibnamefont {Halder}}, \bibinfo {author} {\bibfnamefont
  {K.}~\bibnamefont {Modi}}, \ and\ \bibinfo {author} {\bibfnamefont
  {M.}~\bibnamefont {Paternostro}},\ }\href {\doibase
  10.1103/PhysRevA.93.012305} {\bibfield  {journal} {\bibinfo  {journal} {Phys.
  Rev. A}\ }\textbf {\bibinfo {volume} {93}},\ \bibinfo {pages} {012305}
  (\bibinfo {year} {2016})}\BibitemShut {NoStop}%
\bibitem [{\citenamefont {Fedrizzi}\ \emph {et~al.}(2013)\citenamefont
  {Fedrizzi}, \citenamefont {Zuppardo}, \citenamefont {Gillett}, \citenamefont
  {Broome}, \citenamefont {Almeida}, \citenamefont {Paternostro}, \citenamefont
  {White},\ and\ \citenamefont {Paterek}}]{Fedrizzi2013}%
  \BibitemOpen
  \bibfield  {author} {\bibinfo {author} {\bibfnamefont {A.}~\bibnamefont
  {Fedrizzi}}, \bibinfo {author} {\bibfnamefont {M.}~\bibnamefont {Zuppardo}},
  \bibinfo {author} {\bibfnamefont {G.~G.}\ \bibnamefont {Gillett}}, \bibinfo
  {author} {\bibfnamefont {M.~A.}\ \bibnamefont {Broome}}, \bibinfo {author}
  {\bibfnamefont {M.~P.}\ \bibnamefont {Almeida}}, \bibinfo {author}
  {\bibfnamefont {M.}~\bibnamefont {Paternostro}}, \bibinfo {author}
  {\bibfnamefont {A.~G.}\ \bibnamefont {White}}, \ and\ \bibinfo {author}
  {\bibfnamefont {T.}~\bibnamefont {Paterek}},\ }\href {\doibase
  10.1103/PhysRevLett.111.230504} {\bibfield  {journal} {\bibinfo  {journal}
  {Phys. Rev. Lett.}\ }\textbf {\bibinfo {volume} {111}},\ \bibinfo {pages}
  {230504} (\bibinfo {year} {2013})}\BibitemShut {NoStop}%
\bibitem [{\citenamefont {Vollmer}\ \emph {et~al.}(2013)\citenamefont
  {Vollmer}, \citenamefont {Schulze}, \citenamefont {Eberle}, \citenamefont
  {H\"andchen}, \citenamefont {Fiur\'a\ifmmode~\check{s}\else \v{s}\fi{}ek},\
  and\ \citenamefont {Schnabel}}]{Vollmer2013}%
  \BibitemOpen
  \bibfield  {author} {\bibinfo {author} {\bibfnamefont {C.~E.}\ \bibnamefont
  {Vollmer}}, \bibinfo {author} {\bibfnamefont {D.}~\bibnamefont {Schulze}},
  \bibinfo {author} {\bibfnamefont {T.}~\bibnamefont {Eberle}}, \bibinfo
  {author} {\bibfnamefont {V.}~\bibnamefont {H\"andchen}}, \bibinfo {author}
  {\bibfnamefont {J.}~\bibnamefont {Fiur\'a\ifmmode~\check{s}\else
  \v{s}\fi{}ek}}, \ and\ \bibinfo {author} {\bibfnamefont {R.}~\bibnamefont
  {Schnabel}},\ }\href {\doibase 10.1103/PhysRevLett.111.230505} {\bibfield
  {journal} {\bibinfo  {journal} {Phys. Rev. Lett.}\ }\textbf {\bibinfo
  {volume} {111}},\ \bibinfo {pages} {230505} (\bibinfo {year}
  {2013})}\BibitemShut {NoStop}%
\bibitem [{\citenamefont {Peuntinger}\ \emph {et~al.}(2013)\citenamefont
  {Peuntinger}, \citenamefont {Chille}, \citenamefont
  {Mi\ifmmode~\check{s}\else \v{s}\fi{}ta}, \citenamefont {Korolkova},
  \citenamefont {F\"ortsch}, \citenamefont {Korger}, \citenamefont
  {Marquardt},\ and\ \citenamefont {Leuchs}}]{Peuntinger2013}%
  \BibitemOpen
  \bibfield  {author} {\bibinfo {author} {\bibfnamefont {C.}~\bibnamefont
  {Peuntinger}}, \bibinfo {author} {\bibfnamefont {V.}~\bibnamefont {Chille}},
  \bibinfo {author} {\bibfnamefont {L.}~\bibnamefont {Mi\ifmmode~\check{s}\else
  \v{s}\fi{}ta}}, \bibinfo {author} {\bibfnamefont {N.}~\bibnamefont
  {Korolkova}}, \bibinfo {author} {\bibfnamefont {M.}~\bibnamefont
  {F\"ortsch}}, \bibinfo {author} {\bibfnamefont {J.}~\bibnamefont {Korger}},
  \bibinfo {author} {\bibfnamefont {C.}~\bibnamefont {Marquardt}}, \ and\
  \bibinfo {author} {\bibfnamefont {G.}~\bibnamefont {Leuchs}},\ }\href
  {\doibase 10.1103/PhysRevLett.111.230506} {\bibfield  {journal} {\bibinfo
  {journal} {Phys. Rev. Lett.}\ }\textbf {\bibinfo {volume} {111}},\ \bibinfo
  {pages} {230506} (\bibinfo {year} {2013})}\BibitemShut {NoStop}%
\bibitem [{\citenamefont {Silberhorn}(2013)}]{Silberhorn2013}%
  \BibitemOpen
  \bibfield  {author} {\bibinfo {author} {\bibfnamefont {C.}~\bibnamefont
  {Silberhorn}},\ }\href {\doibase 10.1103/Physics.6.132} {\bibfield  {journal}
  {\bibinfo  {journal} {Physics}\ }\textbf {\bibinfo {volume} {6}},\ \bibinfo
  {pages} {132} (\bibinfo {year} {2013})}\BibitemShut {NoStop}%
\bibitem [{\citenamefont {Werner}(1989)}]{Werner1989}%
  \BibitemOpen
  \bibfield  {author} {\bibinfo {author} {\bibfnamefont {R.~F.}\ \bibnamefont
  {Werner}},\ }\href {\doibase 10.1103/PhysRevA.40.4277} {\bibfield  {journal}
  {\bibinfo  {journal} {Phys. Rev. A}\ }\textbf {\bibinfo {volume} {40}},\
  \bibinfo {pages} {4277} (\bibinfo {year} {1989})}\BibitemShut {NoStop}%
\bibitem [{\citenamefont {Piani}\ \emph {et~al.}(2008)\citenamefont {Piani},
  \citenamefont {Horodecki},\ and\ \citenamefont {Horodecki}}]{Piani2008}%
  \BibitemOpen
  \bibfield  {author} {\bibinfo {author} {\bibfnamefont {M.}~\bibnamefont
  {Piani}}, \bibinfo {author} {\bibfnamefont {P.}~\bibnamefont {Horodecki}}, \
  and\ \bibinfo {author} {\bibfnamefont {R.}~\bibnamefont {Horodecki}},\ }\href
  {\doibase 10.1103/PhysRevLett.100.090502} {\bibfield  {journal} {\bibinfo
  {journal} {Phys. Rev. Lett.}\ }\textbf {\bibinfo {volume} {100}},\ \bibinfo
  {pages} {090502} (\bibinfo {year} {2008})}\BibitemShut {NoStop}%
\bibitem [{\citenamefont {Ferraro}\ \emph {et~al.}(2010)\citenamefont
  {Ferraro}, \citenamefont {Aolita}, \citenamefont {Cavalcanti}, \citenamefont
  {Cucchietti},\ and\ \citenamefont {Ac\'{\i}n}}]{Ferraro2010}%
  \BibitemOpen
  \bibfield  {author} {\bibinfo {author} {\bibfnamefont {A.}~\bibnamefont
  {Ferraro}}, \bibinfo {author} {\bibfnamefont {L.}~\bibnamefont {Aolita}},
  \bibinfo {author} {\bibfnamefont {D.}~\bibnamefont {Cavalcanti}}, \bibinfo
  {author} {\bibfnamefont {F.~M.}\ \bibnamefont {Cucchietti}}, \ and\ \bibinfo
  {author} {\bibfnamefont {A.}~\bibnamefont {Ac\'{\i}n}},\ }\href {\doibase
  10.1103/PhysRevA.81.052318} {\bibfield  {journal} {\bibinfo  {journal} {Phys.
  Rev. A}\ }\textbf {\bibinfo {volume} {81}},\ \bibinfo {pages} {052318}
  (\bibinfo {year} {2010})}\BibitemShut {NoStop}%
\bibitem [{\citenamefont {Bru\ss}(2002)}]{Bruss2002}%
  \BibitemOpen
  \bibfield  {author} {\bibinfo {author} {\bibfnamefont {D.}~\bibnamefont
  {Bru\ss}},\ }\href {\doibase 10.1063/1.1494474} {\bibfield  {journal}
  {\bibinfo  {journal} {J. Math. Phys.}\ }\textbf {\bibinfo {volume} {43}},\
  \bibinfo {pages} {4237} (\bibinfo {year} {2002})}\BibitemShut {NoStop}%
\bibitem [{\citenamefont {{Plenio}}\ and\ \citenamefont
  {{Virmani}}(2007)}]{Plenio2007}%
  \BibitemOpen
  \bibfield  {author} {\bibinfo {author} {\bibfnamefont {M.~B.}\ \bibnamefont
  {{Plenio}}}\ and\ \bibinfo {author} {\bibfnamefont {S.}~\bibnamefont
  {{Virmani}}},\ }\href@noop {} {\bibfield  {journal} {\bibinfo  {journal}
  {Quantum Inf. Comput.}\ }\textbf {\bibinfo {volume} {7}},\ \bibinfo {pages}
  {1} (\bibinfo {year} {2007})},\ \Eprint
  {http://arxiv.org/abs/arXiv:quant-ph/0504163} {arXiv:quant-ph/0504163}
  \BibitemShut {NoStop}%
\bibitem [{\citenamefont {Horodecki}\ \emph {et~al.}(2009)\citenamefont
  {Horodecki}, \citenamefont {Horodecki}, \citenamefont {Horodecki},\ and\
  \citenamefont {Horodecki}}]{Horodecki2009}%
  \BibitemOpen
  \bibfield  {author} {\bibinfo {author} {\bibfnamefont {R.}~\bibnamefont
  {Horodecki}}, \bibinfo {author} {\bibfnamefont {P.}~\bibnamefont
  {Horodecki}}, \bibinfo {author} {\bibfnamefont {M.}~\bibnamefont
  {Horodecki}}, \ and\ \bibinfo {author} {\bibfnamefont {K.}~\bibnamefont
  {Horodecki}},\ }\href {\doibase 10.1103/RevModPhys.81.865} {\bibfield
  {journal} {\bibinfo  {journal} {Rev. Mod. Phys.}\ }\textbf {\bibinfo {volume}
  {81}},\ \bibinfo {pages} {865} (\bibinfo {year} {2009})}\BibitemShut
  {NoStop}%
\bibitem [{\citenamefont {Vedral}\ \emph {et~al.}(1997)\citenamefont {Vedral},
  \citenamefont {Plenio}, \citenamefont {Rippin},\ and\ \citenamefont
  {Knight}}]{Vedral1997}%
  \BibitemOpen
  \bibfield  {author} {\bibinfo {author} {\bibfnamefont {V.}~\bibnamefont
  {Vedral}}, \bibinfo {author} {\bibfnamefont {M.~B.}\ \bibnamefont {Plenio}},
  \bibinfo {author} {\bibfnamefont {M.~A.}\ \bibnamefont {Rippin}}, \ and\
  \bibinfo {author} {\bibfnamefont {P.~L.}\ \bibnamefont {Knight}},\ }\href
  {\doibase 10.1103/PhysRevLett.78.2275} {\bibfield  {journal} {\bibinfo
  {journal} {Phys. Rev. Lett.}\ }\textbf {\bibinfo {volume} {78}},\ \bibinfo
  {pages} {2275} (\bibinfo {year} {1997})}\BibitemShut {NoStop}%
\bibitem [{\citenamefont {Vedral}\ and\ \citenamefont
  {Plenio}(1998)}]{Vedral1998}%
  \BibitemOpen
  \bibfield  {author} {\bibinfo {author} {\bibfnamefont {V.}~\bibnamefont
  {Vedral}}\ and\ \bibinfo {author} {\bibfnamefont {M.~B.}\ \bibnamefont
  {Plenio}},\ }\href {\doibase 10.1103/PhysRevA.57.1619} {\bibfield  {journal}
  {\bibinfo  {journal} {Phys. Rev. A}\ }\textbf {\bibinfo {volume} {57}},\
  \bibinfo {pages} {1619} (\bibinfo {year} {1998})}\BibitemShut {NoStop}%
\bibitem [{Note1()}]{Note1}%
  \BibitemOpen
  \bibinfo {note} {Note that some entanglement measures (like distillable
  entanglement and logarithmic negativity) also vanish for some entangled
  states.}\BibitemShut {Stop}%
\bibitem [{Note2()}]{Note2}%
  \BibitemOpen
  \bibinfo {note} {For a general pure state $\mathinner {|{\psi }\delimiter
  "526930B }$ we will denote the corresponding projector by $\psi $, i.e.,
  $\psi =\mathinner {|{\psi }\delimiter "526930B }\protect \tmspace
  -\thinmuskip {.1667em}\mathinner {\delimiter "426830A {\psi
  }|}$.}\BibitemShut {Stop}%
\bibitem [{\citenamefont {Bennett}\ \emph
  {et~al.}(1996{\natexlab{a}})\citenamefont {Bennett}, \citenamefont
  {Bernstein}, \citenamefont {Popescu},\ and\ \citenamefont
  {Schumacher}}]{Bennett1996b}%
  \BibitemOpen
  \bibfield  {author} {\bibinfo {author} {\bibfnamefont {C.~H.}\ \bibnamefont
  {Bennett}}, \bibinfo {author} {\bibfnamefont {H.~J.}\ \bibnamefont
  {Bernstein}}, \bibinfo {author} {\bibfnamefont {S.}~\bibnamefont {Popescu}},
  \ and\ \bibinfo {author} {\bibfnamefont {B.}~\bibnamefont {Schumacher}},\
  }\href {\doibase 10.1103/PhysRevA.53.2046} {\bibfield  {journal} {\bibinfo
  {journal} {Phys. Rev. A}\ }\textbf {\bibinfo {volume} {53}},\ \bibinfo
  {pages} {2046} (\bibinfo {year} {1996}{\natexlab{a}})}\BibitemShut {NoStop}%
\bibitem [{\citenamefont {Horodecki}\ \emph {et~al.}(1998)\citenamefont
  {Horodecki}, \citenamefont {Horodecki},\ and\ \citenamefont
  {Horodecki}}]{Horodecki1998}%
  \BibitemOpen
  \bibfield  {author} {\bibinfo {author} {\bibfnamefont {M.}~\bibnamefont
  {Horodecki}}, \bibinfo {author} {\bibfnamefont {P.}~\bibnamefont
  {Horodecki}}, \ and\ \bibinfo {author} {\bibfnamefont {R.}~\bibnamefont
  {Horodecki}},\ }\href {\doibase 10.1103/PhysRevLett.80.5239} {\bibfield
  {journal} {\bibinfo  {journal} {Phys. Rev. Lett.}\ }\textbf {\bibinfo
  {volume} {80}},\ \bibinfo {pages} {5239} (\bibinfo {year}
  {1998})}\BibitemShut {NoStop}%
\bibitem [{Note3()}]{Note3}%
  \BibitemOpen
  \bibinfo {note} {Note that $D(\rho ,\sigma )$ does not have to be a distance
  in the mathematical sense, since it does not necessarily fulfill the triangle
  inequality.}\BibitemShut {Stop}%
\bibitem [{\citenamefont {{Uhlmann}}(1998)}]{Uhlmann1998}%
  \BibitemOpen
  \bibfield  {author} {\bibinfo {author} {\bibfnamefont {A.}~\bibnamefont
  {{Uhlmann}}},\ }\href@noop {} {\bibfield  {journal} {\bibinfo  {journal}
  {Open Sys. Inf. Dyn.}\ }\textbf {\bibinfo {volume} {5}},\ \bibinfo {pages}
  {209} (\bibinfo {year} {1998})},\ \Eprint
  {http://arxiv.org/abs/quant-ph/9701014} {quant-ph/9701014} \BibitemShut
  {NoStop}%
\bibitem [{\citenamefont {Bennett}\ \emph
  {et~al.}(1996{\natexlab{b}})\citenamefont {Bennett}, \citenamefont
  {DiVincenzo}, \citenamefont {Smolin},\ and\ \citenamefont
  {Wootters}}]{Bennett1996}%
  \BibitemOpen
  \bibfield  {author} {\bibinfo {author} {\bibfnamefont {C.~H.}\ \bibnamefont
  {Bennett}}, \bibinfo {author} {\bibfnamefont {D.~P.}\ \bibnamefont
  {DiVincenzo}}, \bibinfo {author} {\bibfnamefont {J.~A.}\ \bibnamefont
  {Smolin}}, \ and\ \bibinfo {author} {\bibfnamefont {W.~K.}\ \bibnamefont
  {Wootters}},\ }\href {\doibase 10.1103/PhysRevA.54.3824} {\bibfield
  {journal} {\bibinfo  {journal} {Phys. Rev. A}\ }\textbf {\bibinfo {volume}
  {54}},\ \bibinfo {pages} {3824} (\bibinfo {year}
  {1996}{\natexlab{b}})}\BibitemShut {NoStop}%
\bibitem [{\citenamefont {Horodecki}\ \emph {et~al.}(2000)\citenamefont
  {Horodecki}, \citenamefont {Horodecki},\ and\ \citenamefont
  {Horodecki}}]{Horodecki2000}%
  \BibitemOpen
  \bibfield  {author} {\bibinfo {author} {\bibfnamefont {M.}~\bibnamefont
  {Horodecki}}, \bibinfo {author} {\bibfnamefont {P.}~\bibnamefont
  {Horodecki}}, \ and\ \bibinfo {author} {\bibfnamefont {R.}~\bibnamefont
  {Horodecki}},\ }\href {\doibase 10.1103/PhysRevLett.84.2014} {\bibfield
  {journal} {\bibinfo  {journal} {Phys. Rev. Lett.}\ }\textbf {\bibinfo
  {volume} {84}},\ \bibinfo {pages} {2014} (\bibinfo {year}
  {2000})}\BibitemShut {NoStop}%
\bibitem [{\citenamefont {Hayden}\ \emph {et~al.}(2001)\citenamefont {Hayden},
  \citenamefont {Horodecki},\ and\ \citenamefont {Terhal}}]{Hayden2001}%
  \BibitemOpen
  \bibfield  {author} {\bibinfo {author} {\bibfnamefont {P.~M.}\ \bibnamefont
  {Hayden}}, \bibinfo {author} {\bibfnamefont {M.}~\bibnamefont {Horodecki}}, \
  and\ \bibinfo {author} {\bibfnamefont {B.~M.}\ \bibnamefont {Terhal}},\
  }\href {\doibase 10.1088/0305-4470/34/35/314} {\bibfield  {journal} {\bibinfo
   {journal} {J. Phys. A}\ }\textbf {\bibinfo {volume} {34}},\ \bibinfo {pages}
  {6891} (\bibinfo {year} {2001})}\BibitemShut {NoStop}%
\bibitem [{\citenamefont {Wei}\ and\ \citenamefont {Goldbart}(2003)}]{Wei2003}%
  \BibitemOpen
  \bibfield  {author} {\bibinfo {author} {\bibfnamefont {T.-C.}\ \bibnamefont
  {Wei}}\ and\ \bibinfo {author} {\bibfnamefont {P.~M.}\ \bibnamefont
  {Goldbart}},\ }\href {\doibase 10.1103/PhysRevA.68.042307} {\bibfield
  {journal} {\bibinfo  {journal} {Phys. Rev. A}\ }\textbf {\bibinfo {volume}
  {68}},\ \bibinfo {pages} {042307} (\bibinfo {year} {2003})}\BibitemShut
  {NoStop}%
\bibitem [{\citenamefont {Streltsov}\ \emph {et~al.}(2010)\citenamefont
  {Streltsov}, \citenamefont {Kampermann},\ and\ \citenamefont
  {Bru\ss}}]{Streltsov2010}%
  \BibitemOpen
  \bibfield  {author} {\bibinfo {author} {\bibfnamefont {A.}~\bibnamefont
  {Streltsov}}, \bibinfo {author} {\bibfnamefont {H.}~\bibnamefont
  {Kampermann}}, \ and\ \bibinfo {author} {\bibfnamefont {D.}~\bibnamefont
  {Bru\ss}},\ }\href {\doibase 10.1088/1367-2630/12/12/123004} {\bibfield
  {journal} {\bibinfo  {journal} {New J. Phys.}\ }\textbf {\bibinfo {volume}
  {12}},\ \bibinfo {pages} {123004} (\bibinfo {year} {2010})}\BibitemShut
  {NoStop}%
\bibitem [{\citenamefont {\.{Z}yczkowski}\ \emph {et~al.}(1998)\citenamefont
  {\.{Z}yczkowski}, \citenamefont {Horodecki}, \citenamefont {Sanpera},\ and\
  \citenamefont {Lewenstein}}]{Zyczkowski1998}%
  \BibitemOpen
  \bibfield  {author} {\bibinfo {author} {\bibfnamefont {K.}~\bibnamefont
  {\.{Z}yczkowski}}, \bibinfo {author} {\bibfnamefont {P.}~\bibnamefont
  {Horodecki}}, \bibinfo {author} {\bibfnamefont {A.}~\bibnamefont {Sanpera}},
  \ and\ \bibinfo {author} {\bibfnamefont {M.}~\bibnamefont {Lewenstein}},\
  }\href {\doibase 10.1103/PhysRevA.58.883} {\bibfield  {journal} {\bibinfo
  {journal} {Phys. Rev. A}\ }\textbf {\bibinfo {volume} {58}},\ \bibinfo
  {pages} {883} (\bibinfo {year} {1998})}\BibitemShut {NoStop}%
\bibitem [{\citenamefont {Vidal}\ and\ \citenamefont
  {Werner}(2002)}]{Vidal2002}%
  \BibitemOpen
  \bibfield  {author} {\bibinfo {author} {\bibfnamefont {G.}~\bibnamefont
  {Vidal}}\ and\ \bibinfo {author} {\bibfnamefont {R.~F.}\ \bibnamefont
  {Werner}},\ }\href {\doibase 10.1103/PhysRevA.65.032314} {\bibfield
  {journal} {\bibinfo  {journal} {Phys. Rev. A}\ }\textbf {\bibinfo {volume}
  {65}},\ \bibinfo {pages} {032314} (\bibinfo {year} {2002})}\BibitemShut
  {NoStop}%
\bibitem [{\citenamefont {Horodecki}(1997)}]{Horodecki1997}%
  \BibitemOpen
  \bibfield  {author} {\bibinfo {author} {\bibfnamefont {P.}~\bibnamefont
  {Horodecki}},\ }\href {\doibase 10.1016/S0375-9601(97)00416-7} {\bibfield
  {journal} {\bibinfo  {journal} {Phys. Lett. A}\ }\textbf {\bibinfo {volume}
  {232}},\ \bibinfo {pages} {333 } (\bibinfo {year} {1997})}\BibitemShut
  {NoStop}%
\bibitem [{\citenamefont {Plenio}(2005)}]{Plenio2005}%
  \BibitemOpen
  \bibfield  {author} {\bibinfo {author} {\bibfnamefont {M.~B.}\ \bibnamefont
  {Plenio}},\ }\href {\doibase 10.1103/PhysRevLett.95.090503} {\bibfield
  {journal} {\bibinfo  {journal} {Phys. Rev. Lett.}\ }\textbf {\bibinfo
  {volume} {95}},\ \bibinfo {pages} {090503} (\bibinfo {year}
  {2005})}\BibitemShut {NoStop}%
\bibitem [{\citenamefont {Audenaert}\ \emph {et~al.}(2003)\citenamefont
  {Audenaert}, \citenamefont {Plenio},\ and\ \citenamefont
  {Eisert}}]{Audenaert2003}%
  \BibitemOpen
  \bibfield  {author} {\bibinfo {author} {\bibfnamefont {K.}~\bibnamefont
  {Audenaert}}, \bibinfo {author} {\bibfnamefont {M.~B.}\ \bibnamefont
  {Plenio}}, \ and\ \bibinfo {author} {\bibfnamefont {J.}~\bibnamefont
  {Eisert}},\ }\href {\doibase 10.1103/PhysRevLett.90.027901} {\bibfield
  {journal} {\bibinfo  {journal} {Phys. Rev. Lett.}\ }\textbf {\bibinfo
  {volume} {90}},\ \bibinfo {pages} {027901} (\bibinfo {year}
  {2003})}\BibitemShut {NoStop}%
\bibitem [{\citenamefont {Shor}\ \emph {et~al.}(2001)\citenamefont {Shor},
  \citenamefont {Smolin},\ and\ \citenamefont {Terhal}}]{Shor2001}%
  \BibitemOpen
  \bibfield  {author} {\bibinfo {author} {\bibfnamefont {P.~W.}\ \bibnamefont
  {Shor}}, \bibinfo {author} {\bibfnamefont {J.~A.}\ \bibnamefont {Smolin}}, \
  and\ \bibinfo {author} {\bibfnamefont {B.~M.}\ \bibnamefont {Terhal}},\
  }\href {\doibase 10.1103/PhysRevLett.86.2681} {\bibfield  {journal} {\bibinfo
   {journal} {Phys. Rev. Lett.}\ }\textbf {\bibinfo {volume} {86}},\ \bibinfo
  {pages} {2681} (\bibinfo {year} {2001})}\BibitemShut {NoStop}%
\bibitem [{\citenamefont {Streltsov}\ and\ \citenamefont
  {Zurek}(2013)}]{Streltsov2013}%
  \BibitemOpen
  \bibfield  {author} {\bibinfo {author} {\bibfnamefont {A.}~\bibnamefont
  {Streltsov}}\ and\ \bibinfo {author} {\bibfnamefont {W.~H.}\ \bibnamefont
  {Zurek}},\ }\href {\doibase 10.1103/PhysRevLett.111.040401} {\bibfield
  {journal} {\bibinfo  {journal} {Phys. Rev. Lett.}\ }\textbf {\bibinfo
  {volume} {111}},\ \bibinfo {pages} {040401} (\bibinfo {year}
  {2013})}\BibitemShut {NoStop}%
\bibitem [{\citenamefont {Streltsov}\ \emph
  {et~al.}(2015{\natexlab{b}})\citenamefont {Streltsov}, \citenamefont {Lee},\
  and\ \citenamefont {Adesso}}]{Streltsov2015c}%
  \BibitemOpen
  \bibfield  {author} {\bibinfo {author} {\bibfnamefont {A.}~\bibnamefont
  {Streltsov}}, \bibinfo {author} {\bibfnamefont {S.}~\bibnamefont {Lee}}, \
  and\ \bibinfo {author} {\bibfnamefont {G.}~\bibnamefont {Adesso}},\ }\href
  {\doibase 10.1103/PhysRevLett.115.030505} {\bibfield  {journal} {\bibinfo
  {journal} {Phys. Rev. Lett.}\ }\textbf {\bibinfo {volume} {115}},\ \bibinfo
  {pages} {030505} (\bibinfo {year} {2015}{\natexlab{b}})}\BibitemShut
  {NoStop}%
\bibitem [{Note4()}]{Note4}%
  \BibitemOpen
  \bibinfo {note} {An entanglement breaking channel $\Lambda _{\protect \mathrm
  {eb}}$ has the property that $\Lambda _{\protect \mathrm {eb}}\otimes
  \protect \openone [\rho ]$ is not entangled for any bipartite input state
  $\rho $. We refer to~\cite {Horodecki2003} for more details.}\BibitemShut
  {Stop}%
\bibitem [{\citenamefont {{Datta}}(2010)}]{Datta2010}%
  \BibitemOpen
  \bibfield  {author} {\bibinfo {author} {\bibfnamefont {A.}~\bibnamefont
  {{Datta}}},\ }\href@noop {} {} (\bibinfo {year} {2010}),\ \Eprint
  {http://arxiv.org/abs/1003.5256} {arXiv:1003.5256} \BibitemShut {NoStop}%
\bibitem [{\citenamefont {Koashi}\ and\ \citenamefont
  {Winter}(2004)}]{Koashi2004}%
  \BibitemOpen
  \bibfield  {author} {\bibinfo {author} {\bibfnamefont {M.}~\bibnamefont
  {Koashi}}\ and\ \bibinfo {author} {\bibfnamefont {A.}~\bibnamefont
  {Winter}},\ }\href {\doibase 10.1103/PhysRevA.69.022309} {\bibfield
  {journal} {\bibinfo  {journal} {Phys. Rev. A}\ }\textbf {\bibinfo {volume}
  {69}},\ \bibinfo {pages} {022309} (\bibinfo {year} {2004})}\BibitemShut
  {NoStop}%
\bibitem [{\citenamefont {Fanchini}\ \emph {et~al.}(2011)\citenamefont
  {Fanchini}, \citenamefont {Cornelio}, \citenamefont {de~Oliveira},\ and\
  \citenamefont {Caldeira}}]{Fanchini2011}%
  \BibitemOpen
  \bibfield  {author} {\bibinfo {author} {\bibfnamefont {F.~F.}\ \bibnamefont
  {Fanchini}}, \bibinfo {author} {\bibfnamefont {M.~F.}\ \bibnamefont
  {Cornelio}}, \bibinfo {author} {\bibfnamefont {M.~C.}\ \bibnamefont
  {de~Oliveira}}, \ and\ \bibinfo {author} {\bibfnamefont {A.~O.}\ \bibnamefont
  {Caldeira}},\ }\href {\doibase 10.1103/PhysRevA.84.012313} {\bibfield
  {journal} {\bibinfo  {journal} {Phys. Rev. A}\ }\textbf {\bibinfo {volume}
  {84}},\ \bibinfo {pages} {012313} (\bibinfo {year} {2011})}\BibitemShut
  {NoStop}%
\bibitem [{Note5()}]{Note5}%
  \BibitemOpen
  \bibinfo {note} {Interestingly, Eq.~(\ref {eq:Koashi-Winter}) implies that a
  simple formula for quantum discord for all quantum states is out of reach,
  since such an expression would also allow for an exact evaluation of
  entanglement of formation. Nevertheless, analytical progress on the
  evaluation of discord for particular families of states has been presented
  in~\cite {Ali2010,Ali2010b,Girolami2011}.}\BibitemShut {Stop}%
\bibitem [{Note6()}]{Note6}%
  \BibitemOpen
  \bibinfo {note} {Many authors also consider the minimal distance to the set
  of CQ states, i.e., $\Delta (\rho )=\protect \qopname \relax m{inf}_{\sigma
  \in \protect \mathcal {CQ}}D(\rho ,\sigma )$. We note that $\Delta $ and
  $\protect \mathcal {D}$ coincide for the quantum relative entropy, and
  $\Delta (\rho )\leq \protect \mathcal {D}(\rho )$ in general \cite
  {Modi2012}.}\BibitemShut {Stop}%
\bibitem [{\citenamefont {Modi}\ \emph {et~al.}(2010)\citenamefont {Modi},
  \citenamefont {Paterek}, \citenamefont {Son}, \citenamefont {Vedral},\ and\
  \citenamefont {Williamson}}]{Modi2010}%
  \BibitemOpen
  \bibfield  {author} {\bibinfo {author} {\bibfnamefont {K.}~\bibnamefont
  {Modi}}, \bibinfo {author} {\bibfnamefont {T.}~\bibnamefont {Paterek}},
  \bibinfo {author} {\bibfnamefont {W.}~\bibnamefont {Son}}, \bibinfo {author}
  {\bibfnamefont {V.}~\bibnamefont {Vedral}}, \ and\ \bibinfo {author}
  {\bibfnamefont {M.}~\bibnamefont {Williamson}},\ }\href {\doibase
  10.1103/PhysRevLett.104.080501} {\bibfield  {journal} {\bibinfo  {journal}
  {Phys. Rev. Lett.}\ }\textbf {\bibinfo {volume} {104}},\ \bibinfo {pages}
  {080501} (\bibinfo {year} {2010})}\BibitemShut {NoStop}%
\bibitem [{\citenamefont {Oppenheim}\ \emph {et~al.}(2002)\citenamefont
  {Oppenheim}, \citenamefont {Horodecki}, \citenamefont {Horodecki},\ and\
  \citenamefont {Horodecki}}]{Oppenheim2002}%
  \BibitemOpen
  \bibfield  {author} {\bibinfo {author} {\bibfnamefont {J.}~\bibnamefont
  {Oppenheim}}, \bibinfo {author} {\bibfnamefont {M.}~\bibnamefont
  {Horodecki}}, \bibinfo {author} {\bibfnamefont {P.}~\bibnamefont
  {Horodecki}}, \ and\ \bibinfo {author} {\bibfnamefont {R.}~\bibnamefont
  {Horodecki}},\ }\href {\doibase 10.1103/PhysRevLett.89.180402} {\bibfield
  {journal} {\bibinfo  {journal} {Phys. Rev. Lett.}\ }\textbf {\bibinfo
  {volume} {89}},\ \bibinfo {pages} {180402} (\bibinfo {year}
  {2002})}\BibitemShut {NoStop}%
\bibitem [{\citenamefont {Horodecki}\ \emph {et~al.}(2005)\citenamefont
  {Horodecki}, \citenamefont {Horodecki}, \citenamefont {Horodecki},
  \citenamefont {Oppenheim}, \citenamefont {Sen(De)}, \citenamefont {Sen},\
  and\ \citenamefont {Synak-Radtke}}]{Horodecki2005}%
  \BibitemOpen
  \bibfield  {author} {\bibinfo {author} {\bibfnamefont {M.}~\bibnamefont
  {Horodecki}}, \bibinfo {author} {\bibfnamefont {P.}~\bibnamefont
  {Horodecki}}, \bibinfo {author} {\bibfnamefont {R.}~\bibnamefont
  {Horodecki}}, \bibinfo {author} {\bibfnamefont {J.}~\bibnamefont
  {Oppenheim}}, \bibinfo {author} {\bibfnamefont {A.}~\bibnamefont {Sen(De)}},
  \bibinfo {author} {\bibfnamefont {U.}~\bibnamefont {Sen}}, \ and\ \bibinfo
  {author} {\bibfnamefont {B.}~\bibnamefont {Synak-Radtke}},\ }\href {\doibase
  10.1103/PhysRevA.71.062307} {\bibfield  {journal} {\bibinfo  {journal} {Phys.
  Rev. A}\ }\textbf {\bibinfo {volume} {71}},\ \bibinfo {pages} {062307}
  (\bibinfo {year} {2005})}\BibitemShut {NoStop}%
\bibitem [{\citenamefont {Daki\'{c}}\ \emph {et~al.}(2010)\citenamefont
  {Daki\'{c}}, \citenamefont {Vedral},\ and\ \citenamefont
  {Brukner}}]{Dakic2010}%
  \BibitemOpen
  \bibfield  {author} {\bibinfo {author} {\bibfnamefont {B.}~\bibnamefont
  {Daki\'{c}}}, \bibinfo {author} {\bibfnamefont {V.}~\bibnamefont {Vedral}}, \
  and\ \bibinfo {author} {\bibfnamefont {{\v{C}}.}~\bibnamefont {Brukner}},\
  }\href {\doibase 10.1103/PhysRevLett.105.190502} {\bibfield  {journal}
  {\bibinfo  {journal} {Phys. Rev. Lett.}\ }\textbf {\bibinfo {volume} {105}},\
  \bibinfo {pages} {190502} (\bibinfo {year} {2010})}\BibitemShut {NoStop}%
\bibitem [{\citenamefont {Luo}\ and\ \citenamefont {Fu}(2010)}]{Luo2010}%
  \BibitemOpen
  \bibfield  {author} {\bibinfo {author} {\bibfnamefont {S.}~\bibnamefont
  {Luo}}\ and\ \bibinfo {author} {\bibfnamefont {S.}~\bibnamefont {Fu}},\
  }\href {\doibase 10.1103/PhysRevA.82.034302} {\bibfield  {journal} {\bibinfo
  {journal} {Phys. Rev. A}\ }\textbf {\bibinfo {volume} {82}},\ \bibinfo
  {pages} {034302} (\bibinfo {year} {2010})}\BibitemShut {NoStop}%
\bibitem [{\citenamefont {Piani}(2012)}]{Piani2012}%
  \BibitemOpen
  \bibfield  {author} {\bibinfo {author} {\bibfnamefont {M.}~\bibnamefont
  {Piani}},\ }\href {\doibase 10.1103/PhysRevA.86.034101} {\bibfield  {journal}
  {\bibinfo  {journal} {Phys. Rev. A}\ }\textbf {\bibinfo {volume} {86}},\
  \bibinfo {pages} {034101} (\bibinfo {year} {2012})}\BibitemShut {NoStop}%
\bibitem [{\citenamefont {{Daki{\'c}}}\ \emph {et~al.}(2012)\citenamefont
  {{Daki{\'c}}}, \citenamefont {{Lipp}}, \citenamefont {{Ma}}, \citenamefont
  {{Ringbauer}}, \citenamefont {{Kropatschek}}, \citenamefont {{Barz}},
  \citenamefont {{Paterek}}, \citenamefont {{Vedral}}, \citenamefont
  {{Zeilinger}}, \citenamefont {{Brukner}},\ and\ \citenamefont
  {{Walther}}}]{Dakic2012}%
  \BibitemOpen
  \bibfield  {author} {\bibinfo {author} {\bibfnamefont {B.}~\bibnamefont
  {{Daki{\'c}}}}, \bibinfo {author} {\bibfnamefont {Y.~O.}\ \bibnamefont
  {{Lipp}}}, \bibinfo {author} {\bibfnamefont {X.}~\bibnamefont {{Ma}}},
  \bibinfo {author} {\bibfnamefont {M.}~\bibnamefont {{Ringbauer}}}, \bibinfo
  {author} {\bibfnamefont {S.}~\bibnamefont {{Kropatschek}}}, \bibinfo {author}
  {\bibfnamefont {S.}~\bibnamefont {{Barz}}}, \bibinfo {author} {\bibfnamefont
  {T.}~\bibnamefont {{Paterek}}}, \bibinfo {author} {\bibfnamefont
  {V.}~\bibnamefont {{Vedral}}}, \bibinfo {author} {\bibfnamefont
  {A.}~\bibnamefont {{Zeilinger}}}, \bibinfo {author} {\bibfnamefont {{\v
  C}.}~\bibnamefont {{Brukner}}}, \ and\ \bibinfo {author} {\bibfnamefont
  {P.}~\bibnamefont {{Walther}}},\ }\href {\doibase 10.1038/nphys2377}
  {\bibfield  {journal} {\bibinfo  {journal} {Nature Phys.}\ }\textbf {\bibinfo
  {volume} {8}},\ \bibinfo {pages} {666} (\bibinfo {year} {2012})}\BibitemShut
  {NoStop}%
\bibitem [{\citenamefont {{Brodutch}}\ and\ \citenamefont
  {{Modi}}(2012)}]{Brodutch2011}%
  \BibitemOpen
  \bibfield  {author} {\bibinfo {author} {\bibfnamefont {A.}~\bibnamefont
  {{Brodutch}}}\ and\ \bibinfo {author} {\bibfnamefont {K.}~\bibnamefont
  {{Modi}}},\ }\href@noop {} {\bibfield  {journal} {\bibinfo  {journal}
  {Quantum Inf. Comput.}\ }\textbf {\bibinfo {volume} {12}},\ \bibinfo {pages}
  {0721} (\bibinfo {year} {2012})},\ \Eprint {http://arxiv.org/abs/1108.3649}
  {arXiv:1108.3649} \BibitemShut {NoStop}%
\bibitem [{\citenamefont {Girolami}\ \emph {et~al.}(2014)\citenamefont
  {Girolami}, \citenamefont {Souza}, \citenamefont {Giovannetti}, \citenamefont
  {Tufarelli}, \citenamefont {Filgueiras}, \citenamefont {Sarthour},
  \citenamefont {Soares-Pinto}, \citenamefont {Oliveira},\ and\ \citenamefont
  {Adesso}}]{Girolami2014}%
  \BibitemOpen
  \bibfield  {author} {\bibinfo {author} {\bibfnamefont {D.}~\bibnamefont
  {Girolami}}, \bibinfo {author} {\bibfnamefont {A.~M.}\ \bibnamefont {Souza}},
  \bibinfo {author} {\bibfnamefont {V.}~\bibnamefont {Giovannetti}}, \bibinfo
  {author} {\bibfnamefont {T.}~\bibnamefont {Tufarelli}}, \bibinfo {author}
  {\bibfnamefont {J.~G.}\ \bibnamefont {Filgueiras}}, \bibinfo {author}
  {\bibfnamefont {R.~S.}\ \bibnamefont {Sarthour}}, \bibinfo {author}
  {\bibfnamefont {D.~O.}\ \bibnamefont {Soares-Pinto}}, \bibinfo {author}
  {\bibfnamefont {I.~S.}\ \bibnamefont {Oliveira}}, \ and\ \bibinfo {author}
  {\bibfnamefont {G.}~\bibnamefont {Adesso}},\ }\href {\doibase
  10.1103/PhysRevLett.112.210401} {\bibfield  {journal} {\bibinfo  {journal}
  {Phys. Rev. Lett.}\ }\textbf {\bibinfo {volume} {112}},\ \bibinfo {pages}
  {210401} (\bibinfo {year} {2014})}\BibitemShut {NoStop}%
\bibitem [{\citenamefont {Modi}\ \emph {et~al.}(2011)\citenamefont {Modi},
  \citenamefont {Cable}, \citenamefont {Williamson},\ and\ \citenamefont
  {Vedral}}]{Modi2011}%
  \BibitemOpen
  \bibfield  {author} {\bibinfo {author} {\bibfnamefont {K.}~\bibnamefont
  {Modi}}, \bibinfo {author} {\bibfnamefont {H.}~\bibnamefont {Cable}},
  \bibinfo {author} {\bibfnamefont {M.}~\bibnamefont {Williamson}}, \ and\
  \bibinfo {author} {\bibfnamefont {V.}~\bibnamefont {Vedral}},\ }\href
  {\doibase 10.1103/PhysRevX.1.021022} {\bibfield  {journal} {\bibinfo
  {journal} {Phys. Rev. X}\ }\textbf {\bibinfo {volume} {1}},\ \bibinfo {pages}
  {021022} (\bibinfo {year} {2011})}\BibitemShut {NoStop}%
\bibitem [{\citenamefont {Girolami}\ \emph {et~al.}(2013)\citenamefont
  {Girolami}, \citenamefont {Tufarelli},\ and\ \citenamefont
  {Adesso}}]{Gilolami2013}%
  \BibitemOpen
  \bibfield  {author} {\bibinfo {author} {\bibfnamefont {D.}~\bibnamefont
  {Girolami}}, \bibinfo {author} {\bibfnamefont {T.}~\bibnamefont {Tufarelli}},
  \ and\ \bibinfo {author} {\bibfnamefont {G.}~\bibnamefont {Adesso}},\ }\href
  {\doibase 10.1103/PhysRevLett.110.240402} {\bibfield  {journal} {\bibinfo
  {journal} {Phys. Rev. Lett.}\ }\textbf {\bibinfo {volume} {110}},\ \bibinfo
  {pages} {240402} (\bibinfo {year} {2013})}\BibitemShut {NoStop}%
\bibitem [{\citenamefont {Streltsov}\ \emph
  {et~al.}(2011{\natexlab{a}})\citenamefont {Streltsov}, \citenamefont
  {Kampermann},\ and\ \citenamefont {Bru\ss{}}}]{Streltsov2011b}%
  \BibitemOpen
  \bibfield  {author} {\bibinfo {author} {\bibfnamefont {A.}~\bibnamefont
  {Streltsov}}, \bibinfo {author} {\bibfnamefont {H.}~\bibnamefont
  {Kampermann}}, \ and\ \bibinfo {author} {\bibfnamefont {D.}~\bibnamefont
  {Bru\ss{}}},\ }\href {\doibase 10.1103/PhysRevLett.106.160401} {\bibfield
  {journal} {\bibinfo  {journal} {Phys. Rev. Lett.}\ }\textbf {\bibinfo
  {volume} {106}},\ \bibinfo {pages} {160401} (\bibinfo {year}
  {2011}{\natexlab{a}})}\BibitemShut {NoStop}%
\bibitem [{\citenamefont {Piani}\ \emph {et~al.}(2011)\citenamefont {Piani},
  \citenamefont {Gharibian}, \citenamefont {Adesso}, \citenamefont
  {Calsamiglia}, \citenamefont {Horodecki},\ and\ \citenamefont
  {Winter}}]{Piani2011}%
  \BibitemOpen
  \bibfield  {author} {\bibinfo {author} {\bibfnamefont {M.}~\bibnamefont
  {Piani}}, \bibinfo {author} {\bibfnamefont {S.}~\bibnamefont {Gharibian}},
  \bibinfo {author} {\bibfnamefont {G.}~\bibnamefont {Adesso}}, \bibinfo
  {author} {\bibfnamefont {J.}~\bibnamefont {Calsamiglia}}, \bibinfo {author}
  {\bibfnamefont {P.}~\bibnamefont {Horodecki}}, \ and\ \bibinfo {author}
  {\bibfnamefont {A.}~\bibnamefont {Winter}},\ }\href {\doibase
  10.1103/PhysRevLett.106.220403} {\bibfield  {journal} {\bibinfo  {journal}
  {Phys. Rev. Lett.}\ }\textbf {\bibinfo {volume} {106}},\ \bibinfo {pages}
  {220403} (\bibinfo {year} {2011})}\BibitemShut {NoStop}%
\bibitem [{\citenamefont {Adesso}\ \emph {et~al.}(2014)\citenamefont {Adesso},
  \citenamefont {D'Ambrosio}, \citenamefont {Nagali}, \citenamefont {Piani},\
  and\ \citenamefont {Sciarrino}}]{Adesso2014}%
  \BibitemOpen
  \bibfield  {author} {\bibinfo {author} {\bibfnamefont {G.}~\bibnamefont
  {Adesso}}, \bibinfo {author} {\bibfnamefont {V.}~\bibnamefont {D'Ambrosio}},
  \bibinfo {author} {\bibfnamefont {E.}~\bibnamefont {Nagali}}, \bibinfo
  {author} {\bibfnamefont {M.}~\bibnamefont {Piani}}, \ and\ \bibinfo {author}
  {\bibfnamefont {F.}~\bibnamefont {Sciarrino}},\ }\href {\doibase
  10.1103/PhysRevLett.112.140501} {\bibfield  {journal} {\bibinfo  {journal}
  {Phys. Rev. Lett.}\ }\textbf {\bibinfo {volume} {112}},\ \bibinfo {pages}
  {140501} (\bibinfo {year} {2014})}\BibitemShut {NoStop}%
\bibitem [{\citenamefont {Streltsov}\ \emph
  {et~al.}(2012{\natexlab{b}})\citenamefont {Streltsov}, \citenamefont
  {Adesso}, \citenamefont {Piani},\ and\ \citenamefont
  {Bru\ss{}}}]{Streltsov2012b}%
  \BibitemOpen
  \bibfield  {author} {\bibinfo {author} {\bibfnamefont {A.}~\bibnamefont
  {Streltsov}}, \bibinfo {author} {\bibfnamefont {G.}~\bibnamefont {Adesso}},
  \bibinfo {author} {\bibfnamefont {M.}~\bibnamefont {Piani}}, \ and\ \bibinfo
  {author} {\bibfnamefont {D.}~\bibnamefont {Bru\ss{}}},\ }\href {\doibase
  10.1103/PhysRevLett.109.050503} {\bibfield  {journal} {\bibinfo  {journal}
  {Phys. Rev. Lett.}\ }\textbf {\bibinfo {volume} {109}},\ \bibinfo {pages}
  {050503} (\bibinfo {year} {2012}{\natexlab{b}})}\BibitemShut {NoStop}%
\bibitem [{\citenamefont {{Dhar}}\ \emph {et~al.}(2016)\citenamefont {{Dhar}},
  \citenamefont {{Pal}}, \citenamefont {{Rakshit}}, \citenamefont {{Sen(De)}},\
  and\ \citenamefont {{Sen}}}]{Dhar2016}%
  \BibitemOpen
  \bibfield  {author} {\bibinfo {author} {\bibfnamefont {H.~S.}\ \bibnamefont
  {{Dhar}}}, \bibinfo {author} {\bibfnamefont {A.~K.}\ \bibnamefont {{Pal}}},
  \bibinfo {author} {\bibfnamefont {D.}~\bibnamefont {{Rakshit}}}, \bibinfo
  {author} {\bibfnamefont {A.}~\bibnamefont {{Sen(De)}}}, \ and\ \bibinfo
  {author} {\bibfnamefont {U.}~\bibnamefont {{Sen}}},\ }\href@noop {} {\
  (\bibinfo {year} {2016})},\ \Eprint {http://arxiv.org/abs/1610.01069}
  {arXiv:1610.01069} \BibitemShut {NoStop}%
\bibitem [{\citenamefont {Streltsov}\ \emph
  {et~al.}(2011{\natexlab{b}})\citenamefont {Streltsov}, \citenamefont
  {Kampermann},\ and\ \citenamefont {Bru\ss{}}}]{Streltsov2011}%
  \BibitemOpen
  \bibfield  {author} {\bibinfo {author} {\bibfnamefont {A.}~\bibnamefont
  {Streltsov}}, \bibinfo {author} {\bibfnamefont {H.}~\bibnamefont
  {Kampermann}}, \ and\ \bibinfo {author} {\bibfnamefont {D.}~\bibnamefont
  {Bru\ss{}}},\ }\href {\doibase 10.1103/PhysRevLett.107.170502} {\bibfield
  {journal} {\bibinfo  {journal} {Phys. Rev. Lett.}\ }\textbf {\bibinfo
  {volume} {107}},\ \bibinfo {pages} {170502} (\bibinfo {year}
  {2011}{\natexlab{b}})}\BibitemShut {NoStop}%
\bibitem [{\citenamefont {Fanchini}\ \emph {et~al.}(2010)\citenamefont
  {Fanchini}, \citenamefont {Werlang}, \citenamefont {Brasil}, \citenamefont
  {Arruda},\ and\ \citenamefont {Caldeira}}]{Fanchini2010}%
  \BibitemOpen
  \bibfield  {author} {\bibinfo {author} {\bibfnamefont {F.~F.}\ \bibnamefont
  {Fanchini}}, \bibinfo {author} {\bibfnamefont {T.}~\bibnamefont {Werlang}},
  \bibinfo {author} {\bibfnamefont {C.~A.}\ \bibnamefont {Brasil}}, \bibinfo
  {author} {\bibfnamefont {L.~G.~E.}\ \bibnamefont {Arruda}}, \ and\ \bibinfo
  {author} {\bibfnamefont {A.~O.}\ \bibnamefont {Caldeira}},\ }\href {\doibase
  10.1103/PhysRevA.81.052107} {\bibfield  {journal} {\bibinfo  {journal} {Phys.
  Rev. A}\ }\textbf {\bibinfo {volume} {81}},\ \bibinfo {pages} {052107}
  (\bibinfo {year} {2010})}\BibitemShut {NoStop}%
\bibitem [{\citenamefont {Auccaise}\ \emph {et~al.}(2011)\citenamefont
  {Auccaise}, \citenamefont {Maziero}, \citenamefont {C\'eleri}, \citenamefont
  {Soares-Pinto}, \citenamefont {deAzevedo}, \citenamefont {Bonagamba},
  \citenamefont {Sarthour}, \citenamefont {Oliveira},\ and\ \citenamefont
  {Serra}}]{Auccaise2011}%
  \BibitemOpen
  \bibfield  {author} {\bibinfo {author} {\bibfnamefont {R.}~\bibnamefont
  {Auccaise}}, \bibinfo {author} {\bibfnamefont {J.}~\bibnamefont {Maziero}},
  \bibinfo {author} {\bibfnamefont {L.~C.}\ \bibnamefont {C\'eleri}}, \bibinfo
  {author} {\bibfnamefont {D.~O.}\ \bibnamefont {Soares-Pinto}}, \bibinfo
  {author} {\bibfnamefont {E.~R.}\ \bibnamefont {deAzevedo}}, \bibinfo {author}
  {\bibfnamefont {T.~J.}\ \bibnamefont {Bonagamba}}, \bibinfo {author}
  {\bibfnamefont {R.~S.}\ \bibnamefont {Sarthour}}, \bibinfo {author}
  {\bibfnamefont {I.~S.}\ \bibnamefont {Oliveira}}, \ and\ \bibinfo {author}
  {\bibfnamefont {R.~M.}\ \bibnamefont {Serra}},\ }\href {\doibase
  10.1103/PhysRevLett.107.070501} {\bibfield  {journal} {\bibinfo  {journal}
  {Phys. Rev. Lett.}\ }\textbf {\bibinfo {volume} {107}},\ \bibinfo {pages}
  {070501} (\bibinfo {year} {2011})}\BibitemShut {NoStop}%
\bibitem [{\citenamefont {Girolami}\ and\ \citenamefont
  {Adesso}(2012)}]{Girolami2012}%
  \BibitemOpen
  \bibfield  {author} {\bibinfo {author} {\bibfnamefont {D.}~\bibnamefont
  {Girolami}}\ and\ \bibinfo {author} {\bibfnamefont {G.}~\bibnamefont
  {Adesso}},\ }\href {\doibase 10.1103/PhysRevLett.108.150403} {\bibfield
  {journal} {\bibinfo  {journal} {Phys. Rev. Lett.}\ }\textbf {\bibinfo
  {volume} {108}},\ \bibinfo {pages} {150403} (\bibinfo {year}
  {2012})}\BibitemShut {NoStop}%
\bibitem [{\citenamefont {{Gessner}}\ \emph {et~al.}(2014)\citenamefont
  {{Gessner}}, \citenamefont {{Ramm}}, \citenamefont {{Pruttivarasin}},
  \citenamefont {{Buchleitner}}, \citenamefont {{Breuer}},\ and\ \citenamefont
  {{H{\"a}ffner}}}]{Gessner2014}%
  \BibitemOpen
  \bibfield  {author} {\bibinfo {author} {\bibfnamefont {M.}~\bibnamefont
  {{Gessner}}}, \bibinfo {author} {\bibfnamefont {M.}~\bibnamefont {{Ramm}}},
  \bibinfo {author} {\bibfnamefont {T.}~\bibnamefont {{Pruttivarasin}}},
  \bibinfo {author} {\bibfnamefont {A.}~\bibnamefont {{Buchleitner}}}, \bibinfo
  {author} {\bibfnamefont {H.-P.}\ \bibnamefont {{Breuer}}}, \ and\ \bibinfo
  {author} {\bibfnamefont {H.}~\bibnamefont {{H{\"a}ffner}}},\ }\href {\doibase
  10.1038/nphys2829} {\bibfield  {journal} {\bibinfo  {journal} {Nature Phys.}\
  }\textbf {\bibinfo {volume} {10}},\ \bibinfo {pages} {105} (\bibinfo {year}
  {2014})}\BibitemShut {NoStop}%
\bibitem [{\citenamefont {{Ziman}}\ and\ \citenamefont
  {{Bu\v{z}ek}}(2007)}]{Ziman2007}%
  \BibitemOpen
  \bibfield  {author} {\bibinfo {author} {\bibfnamefont {M.}~\bibnamefont
  {{Ziman}}}\ and\ \bibinfo {author} {\bibfnamefont {V.}~\bibnamefont
  {{Bu\v{z}ek}}},\ }\href@noop {} {} (\bibinfo {year} {2007}),\ \Eprint
  {http://arxiv.org/abs/0707.4401} {arXiv:0707.4401} \BibitemShut {NoStop}%
\bibitem [{\citenamefont {Pal}\ \emph {et~al.}(2014)\citenamefont {Pal},
  \citenamefont {Bandyopadhyay},\ and\ \citenamefont {Ghosh}}]{Pal2014}%
  \BibitemOpen
  \bibfield  {author} {\bibinfo {author} {\bibfnamefont {R.}~\bibnamefont
  {Pal}}, \bibinfo {author} {\bibfnamefont {S.}~\bibnamefont {Bandyopadhyay}},
  \ and\ \bibinfo {author} {\bibfnamefont {S.}~\bibnamefont {Ghosh}},\ }\href
  {\doibase 10.1103/PhysRevA.90.052304} {\bibfield  {journal} {\bibinfo
  {journal} {Phys. Rev. A}\ }\textbf {\bibinfo {volume} {90}},\ \bibinfo
  {pages} {052304} (\bibinfo {year} {2014})}\BibitemShut {NoStop}%
\bibitem [{Note7()}]{Note7}%
  \BibitemOpen
  \bibinfo {note} {The authors of \cite {Pal2014} proved this statement for
  negativity $\protect \mathcal {N}$, which is related to the logarithmic
  negativity as $\protect \mathcal {E}_{\protect \mathrm {n}}=\protect \qopname
  \relax o{log}_{2}(2\protect \mathcal {N}+1)$. Since $\protect \mathcal {N}$
  is a nondecreasing function of $\protect \mathcal {E}_{\protect \mathrm
  {n}}$, it follows that the statement is also true for the logarithmic
  negativity.}\BibitemShut {Stop}%
\bibitem [{\citenamefont {Horodecki}\ \emph {et~al.}(2003)\citenamefont
  {Horodecki}, \citenamefont {Shor},\ and\ \citenamefont
  {Ruskai}}]{Horodecki2003}%
  \BibitemOpen
  \bibfield  {author} {\bibinfo {author} {\bibfnamefont {M.}~\bibnamefont
  {Horodecki}}, \bibinfo {author} {\bibfnamefont {P.~W.}\ \bibnamefont {Shor}},
  \ and\ \bibinfo {author} {\bibfnamefont {M.~B.}\ \bibnamefont {Ruskai}},\
  }\href {\doibase 10.1142/S0129055X03001709} {\bibfield  {journal} {\bibinfo
  {journal} {Rev. Math. Phys.}\ }\textbf {\bibinfo {volume} {15}},\ \bibinfo
  {pages} {629} (\bibinfo {year} {2003})}\BibitemShut {NoStop}%
\bibitem [{\citenamefont {Mi\v{s}ta}\ and\ \citenamefont
  {Korolkova}(2008)}]{Mista2008}%
  \BibitemOpen
  \bibfield  {author} {\bibinfo {author} {\bibfnamefont {L.}~\bibnamefont
  {Mi\v{s}ta}, \bibfnamefont {Jr.}}\ and\ \bibinfo {author} {\bibfnamefont
  {N.}~\bibnamefont {Korolkova}},\ }\href {\doibase 10.1103/PhysRevA.77.050302}
  {\bibfield  {journal} {\bibinfo  {journal} {Phys. Rev. A}\ }\textbf {\bibinfo
  {volume} {77}},\ \bibinfo {pages} {050302} (\bibinfo {year}
  {2008})}\BibitemShut {NoStop}%
\bibitem [{\citenamefont {Baumgratz}\ \emph {et~al.}(2014)\citenamefont
  {Baumgratz}, \citenamefont {Cramer},\ and\ \citenamefont
  {Plenio}}]{Baumgratz2014}%
  \BibitemOpen
  \bibfield  {author} {\bibinfo {author} {\bibfnamefont {T.}~\bibnamefont
  {Baumgratz}}, \bibinfo {author} {\bibfnamefont {M.}~\bibnamefont {Cramer}}, \
  and\ \bibinfo {author} {\bibfnamefont {M.~B.}\ \bibnamefont {Plenio}},\
  }\href {\doibase 10.1103/PhysRevLett.113.140401} {\bibfield  {journal}
  {\bibinfo  {journal} {Phys. Rev. Lett.}\ }\textbf {\bibinfo {volume} {113}},\
  \bibinfo {pages} {140401} (\bibinfo {year} {2014})}\BibitemShut {NoStop}%
\bibitem [{\citenamefont {Winter}\ and\ \citenamefont
  {Yang}(2016)}]{Winter2016}%
  \BibitemOpen
  \bibfield  {author} {\bibinfo {author} {\bibfnamefont {A.}~\bibnamefont
  {Winter}}\ and\ \bibinfo {author} {\bibfnamefont {D.}~\bibnamefont {Yang}},\
  }\href {\doibase 10.1103/PhysRevLett.116.120404} {\bibfield  {journal}
  {\bibinfo  {journal} {Phys. Rev. Lett.}\ }\textbf {\bibinfo {volume} {116}},\
  \bibinfo {pages} {120404} (\bibinfo {year} {2016})}\BibitemShut {NoStop}%
\bibitem [{\citenamefont {{Streltsov}}\ \emph {et~al.}(2016)\citenamefont
  {{Streltsov}}, \citenamefont {{Adesso}},\ and\ \citenamefont
  {{Plenio}}}]{Streltsov2016}%
  \BibitemOpen
  \bibfield  {author} {\bibinfo {author} {\bibfnamefont {A.}~\bibnamefont
  {{Streltsov}}}, \bibinfo {author} {\bibfnamefont {G.}~\bibnamefont
  {{Adesso}}}, \ and\ \bibinfo {author} {\bibfnamefont {M.~B.}\ \bibnamefont
  {{Plenio}}},\ }\href@noop {} {} (\bibinfo {year} {2016}),\ \Eprint
  {http://arxiv.org/abs/1609.02439} {arXiv:1609.02439} \BibitemShut {NoStop}%
\bibitem [{\citenamefont {Bromley}\ \emph {et~al.}(2015)\citenamefont
  {Bromley}, \citenamefont {Cianciaruso},\ and\ \citenamefont
  {Adesso}}]{Bromley2015}%
  \BibitemOpen
  \bibfield  {author} {\bibinfo {author} {\bibfnamefont {T.~R.}\ \bibnamefont
  {Bromley}}, \bibinfo {author} {\bibfnamefont {M.}~\bibnamefont
  {Cianciaruso}}, \ and\ \bibinfo {author} {\bibfnamefont {G.}~\bibnamefont
  {Adesso}},\ }\href {\doibase 10.1103/PhysRevLett.114.210401} {\bibfield
  {journal} {\bibinfo  {journal} {Phys. Rev. Lett.}\ }\textbf {\bibinfo
  {volume} {114}},\ \bibinfo {pages} {210401} (\bibinfo {year}
  {2015})}\BibitemShut {NoStop}%
\bibitem [{\citenamefont {Streltsov}\ \emph
  {et~al.}(2015{\natexlab{c}})\citenamefont {Streltsov}, \citenamefont {Singh},
  \citenamefont {Dhar}, \citenamefont {Bera},\ and\ \citenamefont
  {Adesso}}]{Streltsov2015e}%
  \BibitemOpen
  \bibfield  {author} {\bibinfo {author} {\bibfnamefont {A.}~\bibnamefont
  {Streltsov}}, \bibinfo {author} {\bibfnamefont {U.}~\bibnamefont {Singh}},
  \bibinfo {author} {\bibfnamefont {H.~S.}\ \bibnamefont {Dhar}}, \bibinfo
  {author} {\bibfnamefont {M.~N.}\ \bibnamefont {Bera}}, \ and\ \bibinfo
  {author} {\bibfnamefont {G.}~\bibnamefont {Adesso}},\ }\href {\doibase
  10.1103/PhysRevLett.115.020403} {\bibfield  {journal} {\bibinfo  {journal}
  {Phys. Rev. Lett.}\ }\textbf {\bibinfo {volume} {115}},\ \bibinfo {pages}
  {020403} (\bibinfo {year} {2015}{\natexlab{c}})}\BibitemShut {NoStop}%
\bibitem [{\citenamefont {Chitambar}\ \emph {et~al.}(2016)\citenamefont
  {Chitambar}, \citenamefont {Streltsov}, \citenamefont {Rana}, \citenamefont
  {Bera}, \citenamefont {Adesso},\ and\ \citenamefont
  {Lewenstein}}]{Chitambar2016}%
  \BibitemOpen
  \bibfield  {author} {\bibinfo {author} {\bibfnamefont {E.}~\bibnamefont
  {Chitambar}}, \bibinfo {author} {\bibfnamefont {A.}~\bibnamefont
  {Streltsov}}, \bibinfo {author} {\bibfnamefont {S.}~\bibnamefont {Rana}},
  \bibinfo {author} {\bibfnamefont {M.~N.}\ \bibnamefont {Bera}}, \bibinfo
  {author} {\bibfnamefont {G.}~\bibnamefont {Adesso}}, \ and\ \bibinfo {author}
  {\bibfnamefont {M.}~\bibnamefont {Lewenstein}},\ }\href {\doibase
  10.1103/PhysRevLett.116.070402} {\bibfield  {journal} {\bibinfo  {journal}
  {Phys. Rev. Lett.}\ }\textbf {\bibinfo {volume} {116}},\ \bibinfo {pages}
  {070402} (\bibinfo {year} {2016})}\BibitemShut {NoStop}%
\bibitem [{\citenamefont {Ma}\ \emph {et~al.}(2016)\citenamefont {Ma},
  \citenamefont {Yadin}, \citenamefont {Girolami}, \citenamefont {Vedral},\
  and\ \citenamefont {Gu}}]{Ma2016}%
  \BibitemOpen
  \bibfield  {author} {\bibinfo {author} {\bibfnamefont {J.}~\bibnamefont
  {Ma}}, \bibinfo {author} {\bibfnamefont {B.}~\bibnamefont {Yadin}}, \bibinfo
  {author} {\bibfnamefont {D.}~\bibnamefont {Girolami}}, \bibinfo {author}
  {\bibfnamefont {V.}~\bibnamefont {Vedral}}, \ and\ \bibinfo {author}
  {\bibfnamefont {M.}~\bibnamefont {Gu}},\ }\href {\doibase
  10.1103/PhysRevLett.116.160407} {\bibfield  {journal} {\bibinfo  {journal}
  {Phys. Rev. Lett.}\ }\textbf {\bibinfo {volume} {116}},\ \bibinfo {pages}
  {160407} (\bibinfo {year} {2016})}\BibitemShut {NoStop}%
\bibitem [{\citenamefont {Chitambar}\ and\ \citenamefont
  {Hsieh}(2016)}]{Chitambar2016b}%
  \BibitemOpen
  \bibfield  {author} {\bibinfo {author} {\bibfnamefont {E.}~\bibnamefont
  {Chitambar}}\ and\ \bibinfo {author} {\bibfnamefont {M.-H.}\ \bibnamefont
  {Hsieh}},\ }\href {\doibase 10.1103/PhysRevLett.117.020402} {\bibfield
  {journal} {\bibinfo  {journal} {Phys. Rev. Lett.}\ }\textbf {\bibinfo
  {volume} {117}},\ \bibinfo {pages} {020402} (\bibinfo {year}
  {2016})}\BibitemShut {NoStop}%
\bibitem [{\citenamefont {Matera}\ \emph {et~al.}(2016)\citenamefont {Matera},
  \citenamefont {Egloff}, \citenamefont {Killoran},\ and\ \citenamefont
  {Plenio}}]{Matera2016}%
  \BibitemOpen
  \bibfield  {author} {\bibinfo {author} {\bibfnamefont {J.~M.}\ \bibnamefont
  {Matera}}, \bibinfo {author} {\bibfnamefont {D.}~\bibnamefont {Egloff}},
  \bibinfo {author} {\bibfnamefont {N.}~\bibnamefont {Killoran}}, \ and\
  \bibinfo {author} {\bibfnamefont {M.~B.}\ \bibnamefont {Plenio}},\ }\href
  {\doibase 10.1088/2058-9565/1/1/01LT01} {\bibfield  {journal} {\bibinfo
  {journal} {Quantum Sci. Technol.}\ }\textbf {\bibinfo {volume} {1}},\
  \bibinfo {pages} {01LT01} (\bibinfo {year} {2016})}\BibitemShut {NoStop}%
\bibitem [{\citenamefont {{Streltsov}}\ \emph {et~al.}(2015)\citenamefont
  {{Streltsov}}, \citenamefont {{Rana}}, \citenamefont {{Bera}},\ and\
  \citenamefont {{Lewenstein}}}]{Streltsov2015d}%
  \BibitemOpen
  \bibfield  {author} {\bibinfo {author} {\bibfnamefont {A.}~\bibnamefont
  {{Streltsov}}}, \bibinfo {author} {\bibfnamefont {S.}~\bibnamefont {{Rana}}},
  \bibinfo {author} {\bibfnamefont {M.~N.}\ \bibnamefont {{Bera}}}, \ and\
  \bibinfo {author} {\bibfnamefont {M.}~\bibnamefont {{Lewenstein}}},\
  }\href@noop {} {} (\bibinfo {year} {2015}),\ \Eprint
  {http://arxiv.org/abs/1509.07456} {arXiv:1509.07456} \BibitemShut {NoStop}%
\bibitem [{\citenamefont {{Yadin}}\ \emph {et~al.}(2015)\citenamefont
  {{Yadin}}, \citenamefont {{Ma}}, \citenamefont {{Girolami}}, \citenamefont
  {{Gu}},\ and\ \citenamefont {{Vedral}}}]{Yadin2015}%
  \BibitemOpen
  \bibfield  {author} {\bibinfo {author} {\bibfnamefont {B.}~\bibnamefont
  {{Yadin}}}, \bibinfo {author} {\bibfnamefont {J.}~\bibnamefont {{Ma}}},
  \bibinfo {author} {\bibfnamefont {D.}~\bibnamefont {{Girolami}}}, \bibinfo
  {author} {\bibfnamefont {M.}~\bibnamefont {{Gu}}}, \ and\ \bibinfo {author}
  {\bibfnamefont {V.}~\bibnamefont {{Vedral}}},\ }\href@noop {} {\  (\bibinfo
  {year} {2015})},\ \Eprint {http://arxiv.org/abs/1512.02085}
  {arXiv:1512.02085} \BibitemShut {NoStop}%
\bibitem [{\citenamefont {Ali}\ \emph {et~al.}(2010{\natexlab{a}})\citenamefont
  {Ali}, \citenamefont {Rau},\ and\ \citenamefont {Alber}}]{Ali2010}%
  \BibitemOpen
  \bibfield  {author} {\bibinfo {author} {\bibfnamefont {M.}~\bibnamefont
  {Ali}}, \bibinfo {author} {\bibfnamefont {A.~R.~P.}\ \bibnamefont {Rau}}, \
  and\ \bibinfo {author} {\bibfnamefont {G.}~\bibnamefont {Alber}},\ }\href
  {\doibase 10.1103/PhysRevA.81.042105} {\bibfield  {journal} {\bibinfo
  {journal} {Phys. Rev. A}\ }\textbf {\bibinfo {volume} {81}},\ \bibinfo
  {pages} {042105} (\bibinfo {year} {2010}{\natexlab{a}})}\BibitemShut
  {NoStop}%
\bibitem [{\citenamefont {Ali}\ \emph {et~al.}(2010{\natexlab{b}})\citenamefont
  {Ali}, \citenamefont {Rau},\ and\ \citenamefont {Alber}}]{Ali2010b}%
  \BibitemOpen
  \bibfield  {author} {\bibinfo {author} {\bibfnamefont {M.}~\bibnamefont
  {Ali}}, \bibinfo {author} {\bibfnamefont {A.~R.~P.}\ \bibnamefont {Rau}}, \
  and\ \bibinfo {author} {\bibfnamefont {G.}~\bibnamefont {Alber}},\ }\href
  {\doibase 10.1103/PhysRevA.82.069902} {\bibfield  {journal} {\bibinfo
  {journal} {Phys. Rev. A}\ }\textbf {\bibinfo {volume} {82}},\ \bibinfo
  {pages} {069902} (\bibinfo {year} {2010}{\natexlab{b}})}\BibitemShut
  {NoStop}%
\bibitem [{\citenamefont {Girolami}\ and\ \citenamefont
  {Adesso}(2011)}]{Girolami2011}%
  \BibitemOpen
  \bibfield  {author} {\bibinfo {author} {\bibfnamefont {D.}~\bibnamefont
  {Girolami}}\ and\ \bibinfo {author} {\bibfnamefont {G.}~\bibnamefont
  {Adesso}},\ }\href {\doibase 10.1103/PhysRevA.83.052108} {\bibfield
  {journal} {\bibinfo  {journal} {Phys. Rev. A}\ }\textbf {\bibinfo {volume}
  {83}},\ \bibinfo {pages} {052108} (\bibinfo {year} {2011})}\BibitemShut
  {NoStop}%
\end{thebibliography}%

\end{document}